\documentclass[useAMS]{mn2e}
\usepackage{times,epsfig,amsmath} 
\usepackage{graphicx}
\usepackage{epstopdf}
\usepackage{multirow}

\newcommand{\cxo}{{\it Chandra}}

\newcommand{\xmm}{{\it XMM-Newton}}

\title[The generalized scaling relations]
{The generalized scaling relations for X-ray galaxy clusters: \\ the most powerful mass proxy}

\author[S. Ettori]
{S. Ettori$^{1,2}$ \\
\footnotesize 
 $^1$ INAF, Osservatorio Astronomico di Bologna, via Ranzani 1, I-40127 Bologna, Italy \\
 $^2$ INFN, Sezione di Bologna, viale Berti Pichat 6/2, I-40127 Bologna, Italy \\
 }

\date{Accepted 2013 July 22. Received 2013 July 16; in original form 2013 February 25}

\begin{document}
\maketitle 

\begin{abstract}
The application to observational data of the generalized scaling relations (gSR) presented in Ettori et al. (2012) is here discussed. 
We extend further the formalism of the gSR in the self-similar model for X-ray galaxy clusters, showing that for a generic relation $M_{\rm tot} \propto L^{\alpha} M_g^{\beta} T^{\gamma}$, where $L$, $M_g$ and $T$ are the gas luminosity, mass and temperature, respectively, the values of the slopes lay in the plane $4 \alpha +3 \beta +2 \gamma = 3$. 
Using published dataset, we show that some projections of the gSR are the most efficient relations, holding among observed physical quantities in the X-ray band, to recover the cluster gravitating mass. 
This conclusion is based on the evidence that they provide the lowest $\chi^2$, the lowest total scatter and the lowest intrinsic scatter among the studied scaling laws on both galaxy group and cluster mass scales. 
By the application of the gSR, the intrinsic scatter is reduced in all the cases  down to a relative error on the reconstructed mass below 16 per cent.
The best-fit relations are: $M_{\rm tot} \propto M_g^a T^{1.5-1.5 a}$, with $a \approx 0.4$, and $M_{\rm tot} \propto L^a T^{1.5-2 a}$, with $a \approx 0.15$.
As a by product of this study, we provide the estimates of the gravitating mass at $\Delta=500$ for 120 objects (50 from the Mahdavi et al. 2013 sample, 16 from Maughan 2012; 31 from Pratt et al. 2009; 23 from Sun et al. 2009), 114 of which are unique entries. 
The typical relative error on the mass provided from the gSR only (i.e. not propagating any uncertainty associated with the observed quantities) ranges between 3--5 per cent on cluster scale and is about 10 per cent for galaxy groups.
With respect to the hydrostatic values used to calibrate the gSR, the masses are recovered with deviations in the order of 10 per cent due to the different mix of relaxed/disturbed objects present in the considered samples. In the extreme case of a gSR calibrated with relaxed systems, the hydrostatic mass in disturbed objects is over-estimated by about 20 per cent. 
\end{abstract} 
 

\section{Introduction}

The distribution of the gravitating mass in galaxy cluster is the key ingredient to use them as astrophysical laboratories and cosmological probes.
In the presently favorite hierarchical scenario of cosmic structure formation, direct relations hold between observables in the electromagnetic spectrum and the depth of the cluster potential produced from a matter component expected to be dynamically cold and electromagnetically dark
(see e.g. Allen, Evrard \& Mantz 2011, Kravtsov \& Borgani 2012).

Work in recent years has focused in defining reliable X-ray proxies of the total mass in galaxy clusters.
These X-ray proxies are observables which are at the same time relatively easy to
measure and tightly related to total cluster mass by scaling
relations having low intrinsic scatter as well as a robustly
predicted slope and redshift evolution (e.g. Kravtsov et al. 2006, Maughan 2007 and 2012, Pratt et al. 2009, 
Stanek et al. 2010, Rozo et al. 2009 and 2010, Mantz et al. 2010, Reichert et al. 2011, B\"ohringer et al. 2012; 
see also a recent review in Giodini et al. 2013).    

The X-ray properties of the intra-cluster medium (ICM) are shaped from the evidence that it emits mainly by thermal bremsstrahlung and it is 
hydrostatic equilibrium with the underlying gravitational potential. In this context, the self-similar scenario (e.g. Kaiser 1986, Bryan \& Norman 1998) relates the integrated quantities of the bolometric luminosity, $L$, gas temperature, $T$, and gas mass, $M_g$, to the total mass, $M_{\rm tot}$ in a simple and straightforward way. 
By construction, the cluster mass inside a sphere with volume $4/3 \pi R^3$ corresponding to a mean overdensity $\Delta$ with respect to the critical density of the Universe at the cluster's redshift $z$, $\rho_{c,z}$ , is $M_{\rm tot} = 4/3 \pi \rho_{c,z} \Delta R^3 \propto E_z^2 \Delta R^3$, where $E_z = H_z / H_0 = \left[\Omega_{\rm m} (1+z)^3 + 1 - \Omega_{\rm m} \right]^{1/2}$ represents the cosmic evolution of the Hubble constant $H_0$ for a flat cosmology with matter density parameter $\Omega_{\rm m}$.
From the hydrostatic equilibrium equation (see e.g. Ettori et al. 2013), $M_{\rm tot}$ is directly proportional to the $T R$ or, using the definition above, $E_z M_{\rm tot} \propto T^{3/2}$.
The expression of the bremsstrahlung emissivity $\epsilon \propto \Lambda(T) n_{\rm gas}^2 \propto T^{1/2} n_{\rm gas}^2$ (the
latter relation being valid for systems sufficiently hot, e.g. $> 2$ keV, and assuming a X-ray bolometric emission for which the cooling function $\Lambda(T)$ is $\sim T^{1/2}$) allows us to relate the bolometric luminosity, $L$, and the gas
temperature, $T$: $L \approx \epsilon R^3 \approx T^{1/2} f_{\rm gas}^2 M_{\rm tot}^2 R^{-3} \approx f_{\rm gas}^2 T^2$, 
where we have made use of the above relation between total mass and temperature.
By combining these basic equations, we obtain that the scaling relations
among the X-ray properties and the total mass are (see also Ettori et al. 2004):
$E_z \; M_{\rm tot} \; \propto \; T^{3/2}  \propto \; E_z \;  M_g  \; \propto \; (E_z^{-1} \; L)^{3/4}  \; \propto \;  (E_z Y_X)^{3/5}$.
The latter relation has been introduced from Kravtsov et al. (2006), where the quantity $Y_X = M_g \; T$ is demonstrated to be a very
robust mass proxy being directly proportional to the cluster thermal energy. 
Its scaling relation with $M_{500}$ is characterized by an intrinsic scatter of only 5--7 per cent at fixed $Y_X$, regardless of the dynamical state of the cluster and with a redshift evolution very close to the prediction of self-similar model.  
This robustness of the $M-Y_X$ relation has been studied and confirmed in later work (see, e.g., Arnaud et al. 2007, Maughan 2007, Pratt et al. 2009 on observational data;  Poole et al. 2007, Rasia et al. 2011 and Fabjan et al. 2011 on objects extracted from cosmological hydrodynamical simulations).

The attempt to generalize the simple power-law scaling relations between cluster observables and total mass has become more intensive in the recent past (e.g. Stanek et al. 2010, Okabe et al. 2010, Rozo et al. 2010).

In Ettori et al. (2012; hereafter E12), we have presented new generalized scaling relations with the prospective
to reduce further the scatter between the observed mass proxies and the total cluster mass.
Working on a set of cosmological hydrodynamical simulations, we have found 
a locus of minimum scatter that relates the logarithmic slopes
of the two independent variables considered in that work,
namely the temperature $T$, which traces the depth of the halo gravitational potential, 
and an another observable accounting for distribution of gas density which is more
prone to the affects of the physical processes determining the ICM properties, like
the gas mass $M_g$ or the X-ray luminosity $L$. 
In E12, we show that all the known self-similar scaling laws appear as
particular realizations of generalized scaling relations. 
We predict also the evolution expected for the generalized scaling relations, 
suggesting which relations can be used to maximize the evolutionary effect, for instance to test 
predictions of the self-similar models itself, or, on the contrary, which relations minimize it in the
case of cosmological applications.

In this paper, we present and discuss the application of these 
X-ray {\it generalized scaling relations} on observational data 
to test the improvement introduced from the these relations in reconstructing
the total mass in galaxy clusters. To do this, we do not define any new sample of objects
but use the dataset available in the literature, analyzing them in a homogenous and reproducible way.

The paper is organized as follows. In Section~2, we introduce the 
generalized scaling relations in the context of the self-similar model for X-ray galaxy clusters
and describe how we implement the fit to the selected dataset.  
In Section~3, we discuss the calibration of the generalized scaling relations and 
present the best-fit results in terms of the values of the measured $\chi^2$, 
total and intrinsic scatter.
In Section~4, we summarize our main findings.

\begin{table*}
\caption{Properties of the sample considered in the present analysis: Mahdavi et al. (2013; M13),  Maughan (2012; M12), Pratt et al. (2009; P09) and Sun et al. (2009; S09). 
The median value and the range covered (in brackets) of the listed quantities are shown.}
\begin{tabular}{ccccccc} \hline
Sample & $N$ & $z$ & $M$ & $T$ & $M_g$ & $L$ \\ 
 & & & $10^{14} M_{\odot}$ & keV & $10^{13} M_{\odot}$ & $10^{44}$ erg s$^{-1}$ \\ \hline
M13 & 50 & $0.233 \; (0.152-0.550)$ & $5.7 \; (1.4-13.4)$ & $6.50 \; (3.10-12.10)$ & $8.0 \; (1.4-23.5)$ & $17.0 \; (3.4-131.5)$ \\
M13-CC & 16 & $0.258 \; (0.152-0.464)$ & $4.8 \; (2.1-13.1)$ & $6.20 \; (3.10-12.10)$ & $6.2 \; (2.4-16.3)$ & $19.6 \; (4.9-131.5)$ \\
M13-NCC & 34 & $0.231 \; (0.153-0.550)$ & $5.8 \; (1.4-13.4)$ & $6.80 \; (4.10-11.30)$ & $8.6 \; (1.4-23.5)$ & $16.2 \; (3.4-60.8)$ \\
M12 & 16 & $0.085 \; (0.020-0.230)$ & $3.9 \; (0.8-11.1)$ & $4.67 \; (1.59-8.62)$ & $4.1 \; (0.5-16.2)$ & $ - \; (-)$ \\
P09 & 31 & $0.118 \; (0.056-0.183)$ & $2.6 \; (1.0-7.8)$ & $3.64 \; (2.02-8.24)$ & $3.3 \; (0.8-10.7)$ & $3.8 \; (0.4-36.1)$ \\
S09 & 23 & $0.050 \; (0.012-0.122)$ & $0.8 \; (0.2-1.5)$ & $1.68 \; (0.81-2.68)$ & $0.7 \; (0.1-1.8)$ & $ - \; (-)$ \\
\hline \end{tabular}

\label{tab:prop}
\end{table*}

\section{The generalized scaling laws}

In E12, we have generalized the scaling relations between the total mass $M_{\rm tot}$
and X-ray observables, by considering a more general proxy defined in such a way
that $M_{\rm tot} \propto A^a B^b$, where $A$ is either $M_g$
or $L$ and $B = T$.  
In doing that, we aim to minimize the scatter in the relations between
total mass and observables by (i) relaxing the assumptions done in the
self-similar scenario, (ii) combining information on the depth of the halo gravitational potential
(through the gas temperature $T$) and on the distribution of gas density
(traced by $M_g$ and the X-ray luminosity) that is more
affected by the physical processes determining the ICM global properties,
(iii) adopting a general and flexible function with a
minimal set of free parameters (3 in the general expression -the normalization and the 2 slopes- that
are then reduced to 2 by linking the values of the slopes).

Using a set of cosmological hydrodynamical simulations, we have found 
a locus of minimum scatter that relates the logarithmic slopes $a$ and $b$
of the two independent variables. 
In all cases, this locus is well represented by the lines
$\{A=M_g, \; B=T \} \Rightarrow \; b = -3/2 a +3/2$ and 
$\{A=L, \; B=T \} \Rightarrow \; b = -2 a +3/2$, or, in more concise form,
\begin{equation}
b = 1.5 \; -(1 \;+0.5 d) \; a,
\label{eq:ab}
\end{equation}
where $d$ corresponds to the power to which the gas density appears in the formula of the gas mass ($d = 1$) and luminosity ($d = 2$).
In a similar way, also the evolution with redshift of the total mass can be simply written as $E_z M_{\rm tot} \propto E_z^c$, with
$c = a$ and $-a$ for $A=M_g$ and $L$, respectively, i.e. $c = (3 \, - \, 2 d) \; a$.

The relation in equation~\ref{eq:ab} between the two logarithmic slopes allows us to reduce by one (from 3 to 2)
the number of free parameters in the linear fit of the generalized
scaling law between observables and total mass.

\subsection{The generalized scaling relations in the self-similar model}

The {\it generalized scaling relations} (hereafter gSR) are obtained as the extension of the self-similar model 
when two, or more, observables are used to recover the total gravitating mass.
Indeed, the hydrostatic mass $M_{\rm tot}$ is proportional to $R T$ by definition.
Using $M_g \propto R^3$ implies $M_{\rm tot} \propto M_g^{1/3} T$. 
If we require further that the condition $M_{\rm tot} \propto M_g$ (or $M_{\rm tot} \propto T^{3/2}$) has to be satisfied,
then the relation in equation~\ref{eq:ab} is obtained univocally.

Similarly, we can infer the dependence upon the X-ray bolometric luminosity
($L \propto T^{1/2} M_g^2 R^{-3}$): $M_{\rm tot} \propto R T \propto L^{-1/3} M_g^{2/3} T^{7/6}$ or, equivalently, 
$ \propto L^{-1} M_g^{2} T^{1/2}$. 
Then, we can solve for any combination of observables to recover the relation in  equation~\ref{eq:ab}.

It is worth noticing that these observables ($L, M_g, T$) are the only ones accessible directly through the X-ray analysis:
the luminosity is provided from the observed count rate once a thermal model and redshift are assumed; 
the gas mass is obtained as integral of the modelled (or deprojected) X-ray surface brightness;
the gas temperature is constrained from the continuum of the spectral thermal model.

More generally, we can write
\begin{equation}
M_{\rm tot} \propto L^{\alpha} M_g^{\beta} T^{\gamma}
\label{eq:gen}
\end{equation}
with the exponents
$(\alpha, \beta, \gamma)$ satisfying, in the self-similar scenario, the equation
\begin{equation}
4 \alpha + 3 \beta + 2 \gamma = 3.
\label{eq:plane}
\end{equation}
The projections of this plane in the cartesian axes $(\alpha, \beta, \gamma)$ provide the subset of relations discussed in E12:
\begin{align}
(\alpha = 0) & \;  \gamma = 3/2 \; -3/2 \; \beta    \nonumber \\
(\beta = 0) & \;  \gamma = 3/2 \; -2 \; \alpha \nonumber \\
(\gamma = 0) & \;  \beta = 1 \; -4/3 \; \alpha.  
\label{eq:proj}
\end{align}

The self-similar evolution of the equation~\ref{eq:gen} is then $E_z M_{\rm tot} \propto (E_z^{-1} L)^{\alpha} \;
(E_z M_g)^{\beta} T^{\gamma} \sim E_z^{\epsilon}$, with
$\epsilon = -\alpha +\beta$.

It is worth noticing that these gSRs reduce to the standard self-similar laws with a single observables 
for a proper value of the slope of equation~\ref{eq:gen} (or equation~\ref{eq:proj}):
one recovers $M_{\rm tot} \propto T^{3/2}$ with $(\alpha, \beta) = (0, 0)$; 
$M_{\rm tot} \propto M_g$ with $(\alpha, \gamma) = (0, 0)$;
$M_{\rm tot} \propto Y_X^{3/5}$ with $(\alpha, \beta) = (0, 3/5)$;
$M_{\rm tot} \propto L^{3/4}$ with $(\beta, \gamma) = (0, 0)$;
$M_{\rm tot} \propto \left(LT\right)^{1/2}$, which is the relation corresponding to
$M_{\rm tot} \propto Y_X^{3/5}$ once gas mass is replaced by luminosity,
fixing $(\beta, \gamma) = (0, 1/2)$.

In the following analysis, we investigate particularly some projections of the gSR in equation~\ref{eq:plane}, focusing our analysis on those relations that minimize the scatter in recovering the total cluster gravitating mass.

\begin{table}
\caption{Scatter and $\chi^2$ measured in the listed scaling relations by using the data quoted in Mahdavi et al. (2013; M13),  Maughan (2012; M12), Pratt et al. (2009; P09) and Sun et al. (2009; S09). The degrees-of-freedom $D$ is the number of objects in the sample minus 2, the number of free parameters ($n, a$) used in the fit.}
\begin{tabular}{ccccc} \hline
relation & $D$ & $\chi^2$ & $\sigma_M$ & $\sigma_I$ \\ \hline
$M - T$ (M13) & 48 & $139.8$ & $0.122$ & $0.085^{+0.017}_{-0.013}$ \\
$M - M_g$ (M13) & 48 & $210.3$ & $0.131$ & $0.101^{+0.018}_{-0.013}$ \\
$M - L$ (M13) & 48 & $548.5$ & $0.157$ & $0.136^{+0.020}_{-0.014}$ \\
$\boldsymbol{ M - M_g \; T}$ (M13) & 48 & $124.6$ & $0.108$ & $0.071^{+0.015}_{-0.011}$ \\
$\boldsymbol{ M - L \; T}$ (M13) & 48 & $123.7$ & $0.113$ & $0.074^{+0.015}_{-0.012}$ \\
$\boldsymbol{ M - L \; M_g}$ (M13) & 48 & $213.6$ & $0.129$ & $0.100^{+0.018}_{-0.013}$ \\
\hline 
$M - T$ (M13-CC) & 14 & $44.2$ & $0.098$ & $0.070^{+0.030}_{-0.017}$ \\
$M - M_g$ (M13-CC) & 14 & $58.3$ & $0.127$ & $0.102^{+0.038}_{-0.023}$ \\
$M - L$ (M13-CC) & 14 & $160.7$ & $0.129$ & $0.113^{+0.035}_{-0.019}$ \\
$\boldsymbol{ M - M_g \; T}$ (M13-CC) & 14 & $43.4$ & $0.084$ & $0.060^{+0.026}_{-0.015}$ \\
$\boldsymbol{ M - L \; T}$ (M13-CC) & 14 & $40.3$ & $0.079$ & $0.056^{+0.025}_{-0.014}$ \\
$\boldsymbol{ M - L \; M_g}$ (M13-CC) & 14 & $58.7$ & $0.124$ & $0.098^{+0.038}_{-0.022}$ \\
\hline 
$M - T$ (M12) & 14 & $18.1$ & $0.064$ & $0.025^{+0.025}_{-0.025}$ \\
$M - M_g$ (M12) & 14 & $13.1$ & $0.065$ & $0.000^{+0.038}_{-0.000}$ \\
$\boldsymbol{ M - M_g \; T}$ (M12) & 14 & $14.0$ & $0.052$ & $0.002^{+0.034}_{-0.002}$ \\
\hline 
$M - T$ (P09) & 29 & $164.9$ & $0.055$ & $0.045^{+0.010}_{-0.007}$ \\
$M - M_g$ (P09) & 29 & $62.5$ & $0.029$ & $0.017^{+0.006}_{-0.004}$ \\
$M - L$ (P09) & 29 & $3794.0$ & $0.085$ & $0.085^{+0.013}_{-0.009}$ \\
$\boldsymbol{ M - M_g \; T}$ (P09) & 29 & $48.9$ & $0.018$ & $0.010^{+0.003}_{-0.003}$ \\
$\boldsymbol{ M - L \; T}$ (P09) & 29 & $153.1$ & $0.047$ & $0.038^{+0.008}_{-0.005}$ \\
$\boldsymbol{ M - L \; M_g}$ (P09) & 29 & $183.1$ & $0.068$ & $0.059^{+0.012}_{-0.008}$ \\
\hline 
$M - T$ (S09) & 21 & $15.8$ & $0.081$ & $0.000^{+0.022}_{-0.000}$ \\
$M - M_g$ (S09) & 21 & $13.6$ & $0.075$ & $0.000$ -- \\
$M - L$ (S09) & 21 & $10.1$ & $0.060$ & $0.000$ -- \\
\hline 
\end{tabular}

\label{tab:bestfit_scat}
\end{table}

\begin{table}
\caption{Best-fit parameters of the generalized scaling laws. The errors on $n$ and $a$ are used in combination with the element off-diagonal ($cov_{na}$) of the covariance matrix of the fit to evaluate the error on $M_{fit}$ through a standard error propagation (see equation~\ref{eq:errm}).
}
\begin{tabular}{cccc} \hline
Relation & $n$ & $a$ & $cov_{na}$ \\ \hline
\multicolumn{4}{c}{$\boldsymbol{ M - T}$} \\
M13 & $-0.074 \pm 0.013$ & $1.689 \pm 0.071$ & $-6.234 \times 10^{-4}$ \\
M13-CC & $-0.019 \pm 0.015$ & $1.714 \pm 0.105$ & $-8.796 \times 10^{-4}$ \\
M12 & $-0.059 \pm 0.012$ & $1.643 \pm 0.060$ & $-2.356 \times 10^{-5}$ \\
P09 & $-0.065 \pm 0.004$ & $1.572 \pm 0.021$ & $3.122 \times 10^{-5}$ \\
S09 & $-0.042 \pm 0.064$ & $1.751 \pm 0.122$ & $7.506 \times 10^{-3}$ \\
\multicolumn{4}{c}{$\boldsymbol{ M - M_g}$} \\
M13 & $-0.102 \pm 0.010$ & $0.872 \pm 0.027$ & $-2.044 \times 10^{-4}$ \\
M13-CC & $-0.143 \pm 0.019$ & $1.036 \pm 0.059$ & $-8.949 \times 10^{-4}$ \\
M12 & $-0.046 \pm 0.014$ & $0.815 \pm 0.033$ & $2.817 \times 10^{-5}$ \\
P09 & $-0.101 \pm 0.003$ & $0.839 \pm 0.009$ & $8.056 \times 10^{-6}$ \\
S09 & $-0.056 \pm 0.068$ & $0.865 \pm 0.067$ & $4.378 \times 10^{-3}$ \\
\multicolumn{4}{c}{$\boldsymbol{ M - L}$} \\
M13 & $0.012 \pm 0.008$ & $0.491 \pm 0.019$ & $-8.761 \times 10^{-5}$ \\
M13-CC & $-0.064 \pm 0.016$ & $0.469 \pm 0.031$ & $-3.709 \times 10^{-4}$ \\
P09 & $-0.030 \pm 0.002$ & $0.497 \pm 0.003$ & $2.266 \times 10^{-6}$ \\
\multicolumn{4}{c}{$\boldsymbol{ M - M_g \; T}$} \\
M13 & $-0.095 \pm 0.009$ & $0.417 \pm 0.061$ & $-3.355 \times 10^{-4}$ \\
M13-CC & $-0.049 \pm 0.021$ & $0.349 \pm 0.135$ & $-2.381 \times 10^{-3}$ \\
M12 & $-0.054 \pm 0.012$ & $0.348 \pm 0.104$ & $1.652 \times 10^{-4}$ \\
P09 & $-0.077 \pm 0.003$ & $0.419 \pm 0.027$ & $-1.223 \times 10^{-5}$ \\
S09 & $-0.045 \pm 0.040$ & $0.500 \pm 0.147$ & $5.297 \times 10^{-3}$ \\
\multicolumn{4}{c}{$\boldsymbol{ M - L \; T}$} \\
M13 & $-0.064 \pm 0.008$ & $0.159 \pm 0.030$ & $-5.681 \times 10^{-5}$ \\
M13-CC & $-0.036 \pm 0.016$ & $0.141 \pm 0.047$ & $-5.254 \times 10^{-4}$ \\
P09 & $-0.058 \pm 0.004$ & $0.073 \pm 0.014$ & $2.987 \times 10^{-5}$ \\
\multicolumn{4}{c}{$\boldsymbol{ M - L \; M_g}$} \\
M13 & $-0.125 \pm 0.007$ & $0.119 \pm 0.027$ & $7.559 \times 10^{-5}$ \\
M13-CC & $-0.134 \pm 0.012$ & $-0.007 \pm 0.058$ & $-2.245 \times 10^{-4}$ \\
P09 & $-0.123 \pm 0.006$ & $-0.203 \pm 0.024$ & $1.102 \times 10^{-4}$ \\
\hline \end{tabular}

\label{tab:bestfit_par}
\end{table}

\subsection{Fitting the scaling relations}

In this work, we want to compare how the assumed linear relation between logarithmic values of the observed quantities and of the gravitational mass determined through the equation of the hydrostatic equilibrium, $M_{\rm tot} \equiv M_{\rm HSE} \equiv M$, performs and, in particular, which is the combination of observables that minimizes the scatter in reconstructing the galaxy cluster mass. 
Among the relations satisfying equation~\ref{eq:plane}, we focus on the most promising for our goal, $M \propto M_g T$ and $M \propto L T$, that are obtained by requiring $\alpha=0$ and $\beta=0$, respectively. 

Operationally, we adopt the following procedure.
We build the variables
\begin{align}
\mathcal{Y} = & \log\left( \frac{E_z \; M_{\rm HSE}}{5 \times 10^{14} M_{\odot}} \right)  \nonumber \\
\mathcal{A} = & \log(A);  \;  A = {\rm either} \; \frac{E_z \; M_{g}}{5 \times 10^{13} M_{\odot}}
  \; {\rm or} \; \frac{E_z^{-1} \;  L_{\rm bol}}{10^{45} {\rm erg \, s}^{-1}} \nonumber \\
\mathcal{B} = & \log(B); \;  B = \frac{T}{5 {\rm keV}}
\label{eq:log}
\end{align}
where ``$\log$'' indicates the base-10 logarithm, 
and we consider an associated error obtained through the propagation of the measured uncertainties.

Then, we fit the linear function $\mathcal{Y} = n \; +a \mathcal{A} \; +b \mathcal{B}$.
The best-fit parameters are obtained by minimizing the merit function:
\begin{align}
 \chi^2 = & \sum_{i=1}^N \frac{(\mathcal{Y}_i \; -n \; -a \mathcal{A}_i \; -b \mathcal{B}_i)^2}{ \epsilon_i^2}  \nonumber \\
 \epsilon_i^2 = & \epsilon_{\mathcal{Y}, i}^2 \; +a^2 \epsilon_{\mathcal{A}, i}^2 \; +b^2 \epsilon_{\mathcal{B}, i}^2
 \label{eq:chi2}
\end{align}
where $b$ is related to $a$ through equation~\ref{eq:ab}, $N$ is the number of data points and $D=N-2$ are the degrees of freedom.

The fit is performed using the {\tt IDL} routine {\it MPFIT} (Markwardt 2008).

To evaluate further the performance of the gSR with respect to the standard scaling laws, we have also estimated 
the total and the intrinsic scatter.

Here, we define the {\it total scatter} on the logarithmic value of the mass $\sigma_M$ as the sum, divided by the degrees-of-freedom, of the residuals of the observed measurements with respect to the best-fit line:
\begin{align}
w_i = & {\rm either} \; \frac{N / \epsilon^2_i}{\sum_{j=1}^N 1/\epsilon^2_j} \; {\rm or} \; 1 \nonumber \\
\sigma_M^2 = & \frac{1}{D} \sum_{i=1}^N w_i (\mathcal{Y}_i \; -n \; -a \mathcal{A}_i \; -b \mathcal{B}_i)^2.
\end{align}
The two definitions of the weights $w_i$ do not change significantly the measured scatter. Hereafter, we define $w_i = 1$.

The {\it intrinsic scatter} is a constant value $\sigma_I$ that is determined by adding it in quadrature to $\epsilon_i$ in equation~\ref{eq:chi2}, once the minimum $\chi^2$ is estimated, and looking for the values that satisfy the relation
\begin{equation}
\chi^2_{\rm red} = \frac{\chi^2}{D} = 1 \pm \sqrt{ \frac{2}{D} },
\label{eq:scat_int}
\end{equation}
where the dispersion around 1 of the reduced $\chi^2$, $\chi^2_{\rm red}$, is strictly valid in the limit of large $D$.

By construction, the intrinsic scatter estimated through equation~\ref{eq:scat_int} translates then in a contribution (to be added in quadrature) to the
relative error on the mass equals to $\ln(10) \, \sigma_I \approx 2.30 \, \sigma_I$.

\begin{table*}
\caption{The values of (1st row) mean and median and (2nd row, inside square brackets) dispersion and Inter-Quartile-Range (IQR $\approx 1.35 \sigma$ are quoted for the ratios $M_{fit}/M_{\rm HSE}$, where $M_{fit}$ are evaluated according to the sample and the generalized scaling law shown in the first row and $M_{\rm HSE}$ is the hydrostatic mass value for the objects in the sample indicated in the first column (e.g.: using the $M_g T$ generalized scaling relation calibrated with the objects in the M13 sample --1st column-- we are able to reconstruct the total masses in the M12 sample --4th row-- with an average/median ratio $M_{fit}/M_{\rm HSE}$ of 0.905/0.915).
}
\begin{tabular}{c|cccccccc} \hline
Sample & M13$- M_g \, T$ & M13-CC$- M_g \, T$ & M12$- M_g \, T$ & S09$- M_g \, T$ & M13$- L \, T$ & M13-CC$- L \, T$ \\ \hline
\multirow{2}{*}{M13} & $1.038, 1.014$ & $1.135, 1.109$ & $1.124, 1.099$ & $1.185, 1.143$ & $0.998, 0.980$ & $1.066, 1.050$ \\
 & $(0.261, 0.303)$ & $(0.284, 0.321)$ & $(0.281, 0.318)$ & $(0.305, 0.349)$ & $(0.265, 0.257)$ & $(0.285, 0.275)$ \\
 \\
\multirow{2}{*}{M13-CC} & $0.915, 0.937$ & $0.998, 1.030$ & $0.988, 1.020$ & $1.046, 1.050$ & $0.939, 0.914$ & $0.994, 0.969$ \\
 & $(0.163, 0.238)$ & $(0.174, 0.279)$ & $(0.172, 0.277)$ & $(0.194, 0.255)$ & $(0.160, 0.203)$ & $(0.168, 0.210)$ \\
 \\
\multirow{2}{*}{M13-NCC} & $1.097, 1.047$ & $1.200, 1.129$ & $1.188, 1.118$ & $1.251, 1.194$ & $1.026, 1.020$ & $1.100, 1.092$ \\
 & $(0.280, 0.244)$ & $(0.305, 0.244)$ & $(0.302, 0.242)$ & $(0.327, 0.339)$ & $(0.300, 0.263)$ & $(0.322, 0.287)$ \\
 \\
\multirow{2}{*}{M12} & $0.905, 0.915$ & $1.014, 1.040$ & $1.005, 1.031$ & $1.002, 1.051$ & $-, -$ & $-, -$ \\
 & $(0.109, 0.170)$ & $(0.116, 0.207)$ & $(0.115, 0.206)$ & $(0.135, 0.191)$ & $(-, -)$ & $(-, -)$ \\
 \\
\multirow{2}{*}{P09} & $0.962, 0.954$ & $1.072, 1.070$ & $1.062, 1.061$ & $1.072, 1.073$ & $0.960, 0.956$ & $1.035, 1.018$ \\
 & $(0.040, 0.063)$ & $(0.052, 0.077)$ & $(0.052, 0.077)$ & $(0.047, 0.080)$ & $(0.095, 0.097)$ & $(0.103, 0.105)$ \\
 \\
\multirow{2}{*}{S09} & $0.925, 0.940$ & $1.064, 1.088$ & $1.055, 1.077$ & $0.994, 0.956$ & $-, -$ & $-, -$ \\
 & $(0.130, 0.146)$ & $(0.154, 0.197)$ & $(0.153, 0.197)$ & $(0.139, 0.170)$ & $(-, -)$ & $(-, -)$ \\
 \\
\hline \end{tabular}

\label{tab:bestfit_ratio}
\end{table*}

\begin{figure*}
\hbox{
 \includegraphics[width=0.5\textwidth, keepaspectratio]{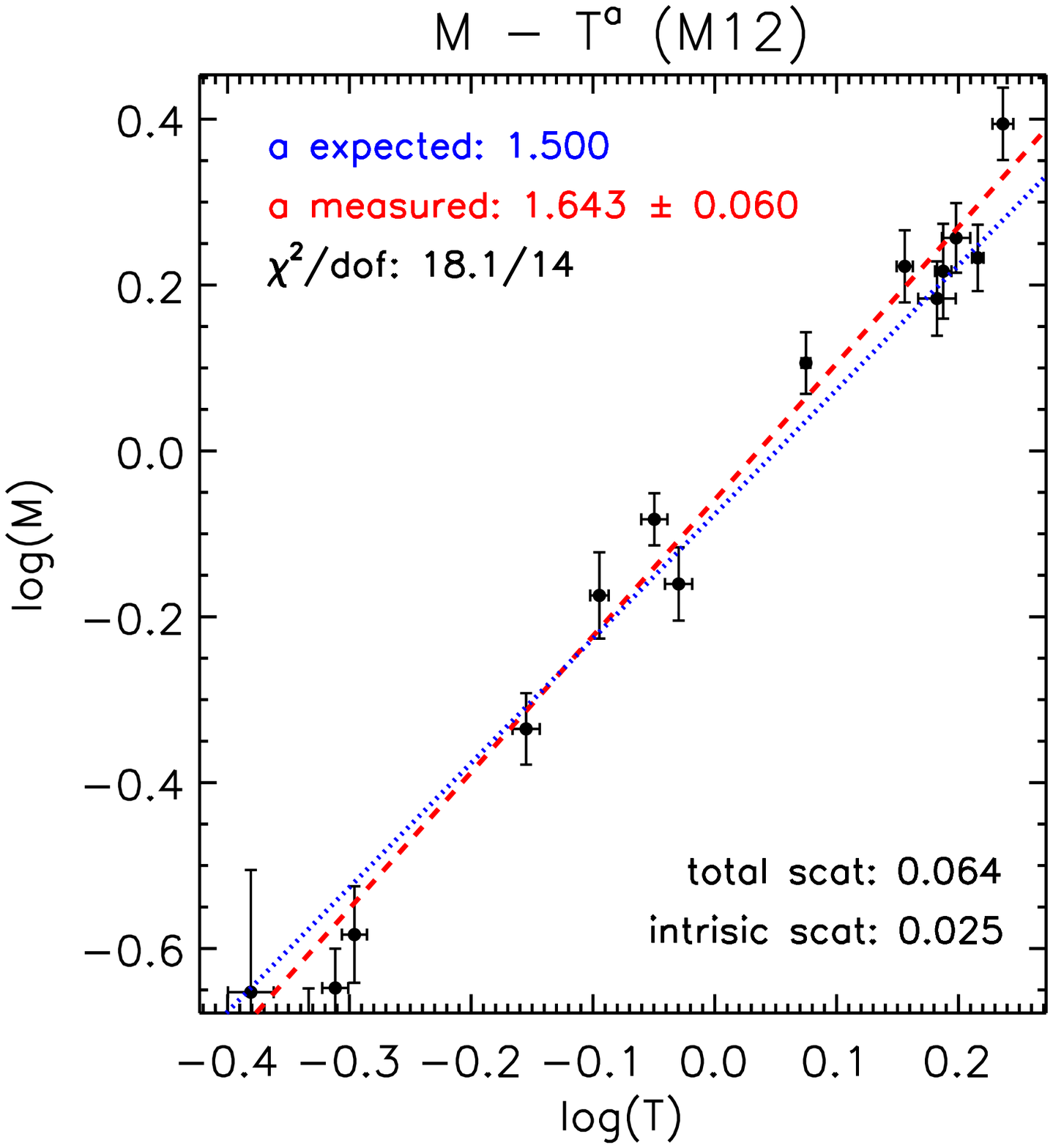}
  \includegraphics[width=0.5\textwidth, keepaspectratio]{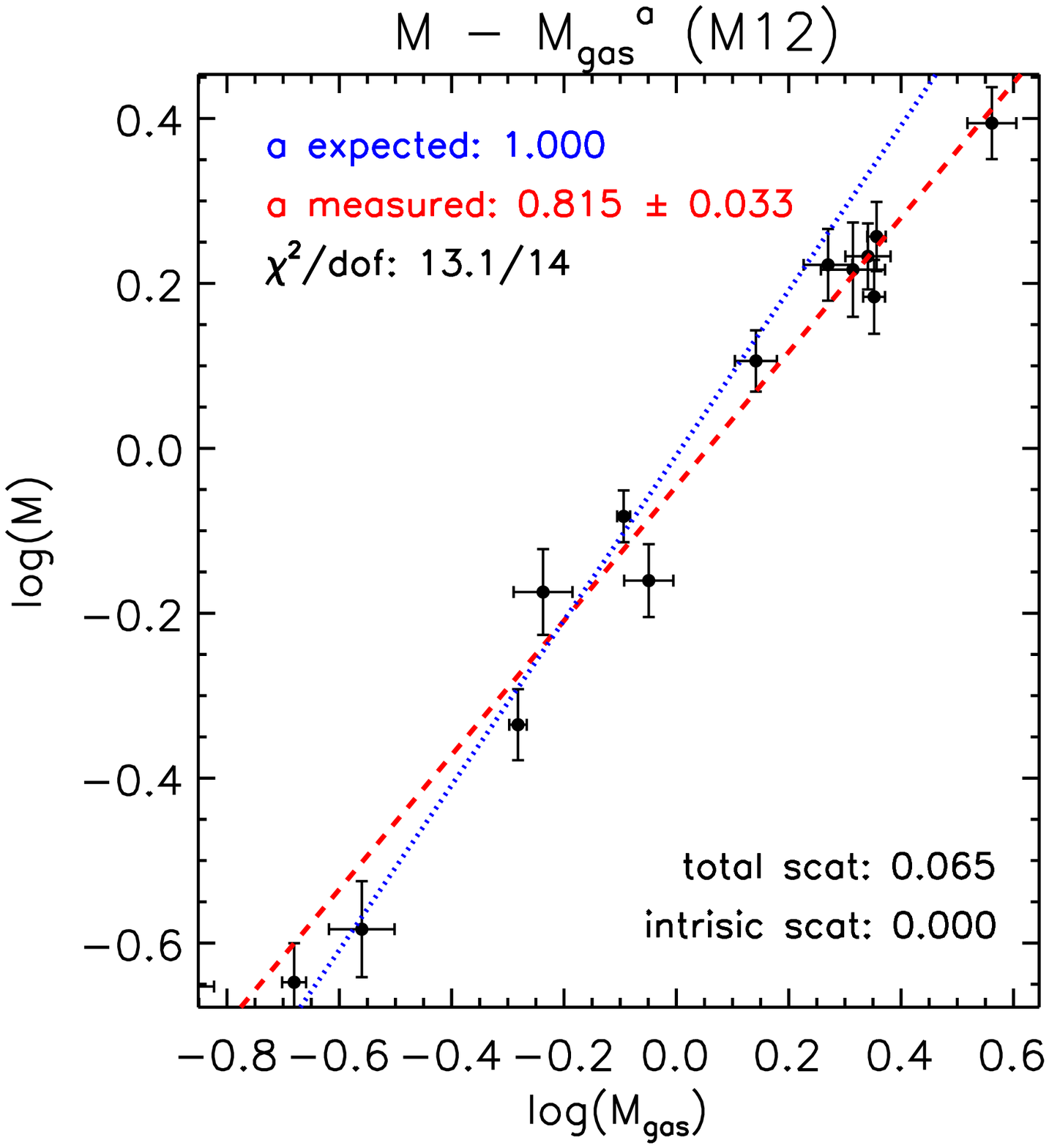}
} \hbox{
  \includegraphics[width=0.5\textwidth, keepaspectratio]{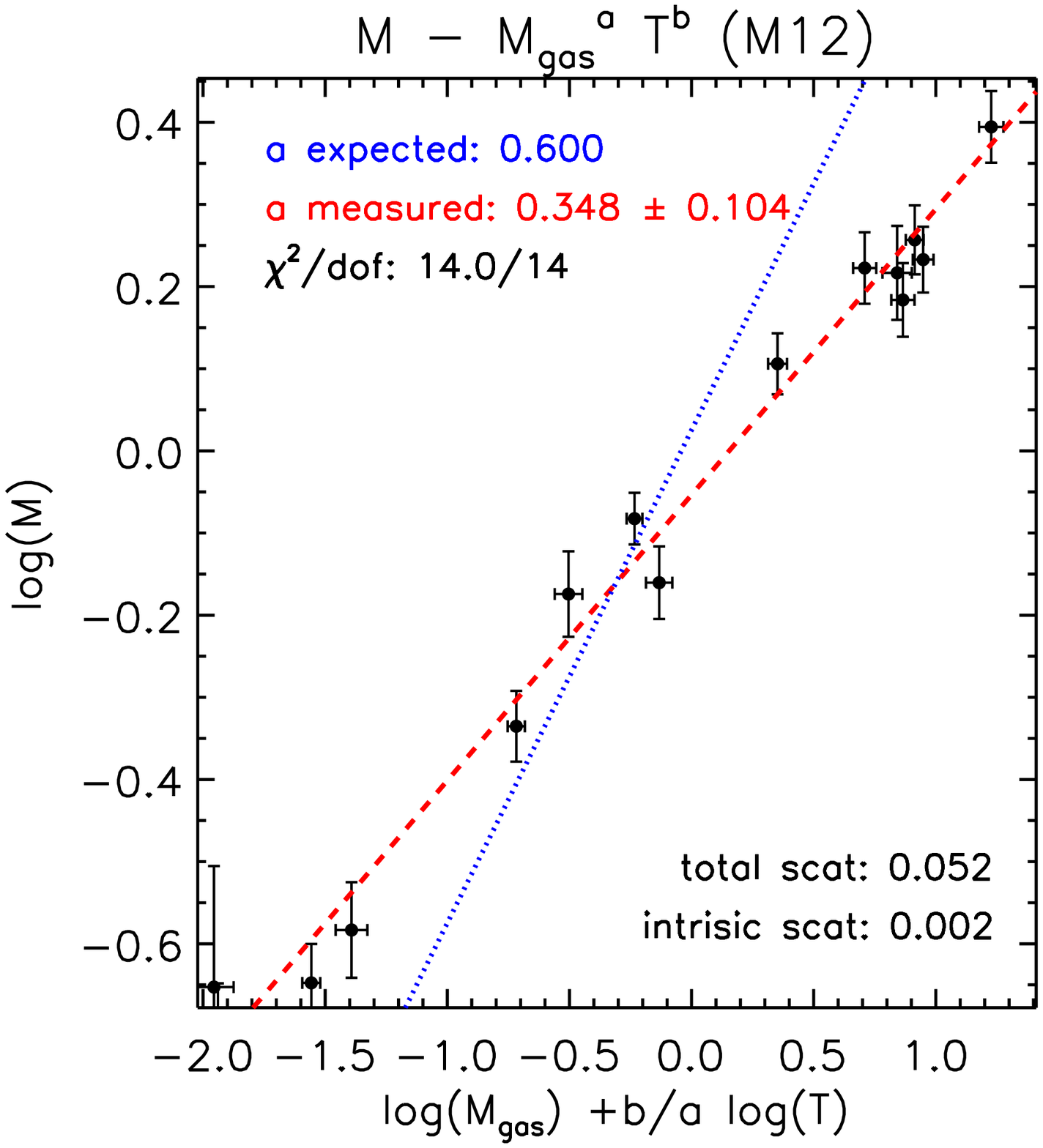}
  \includegraphics[width=0.5\textwidth, keepaspectratio]{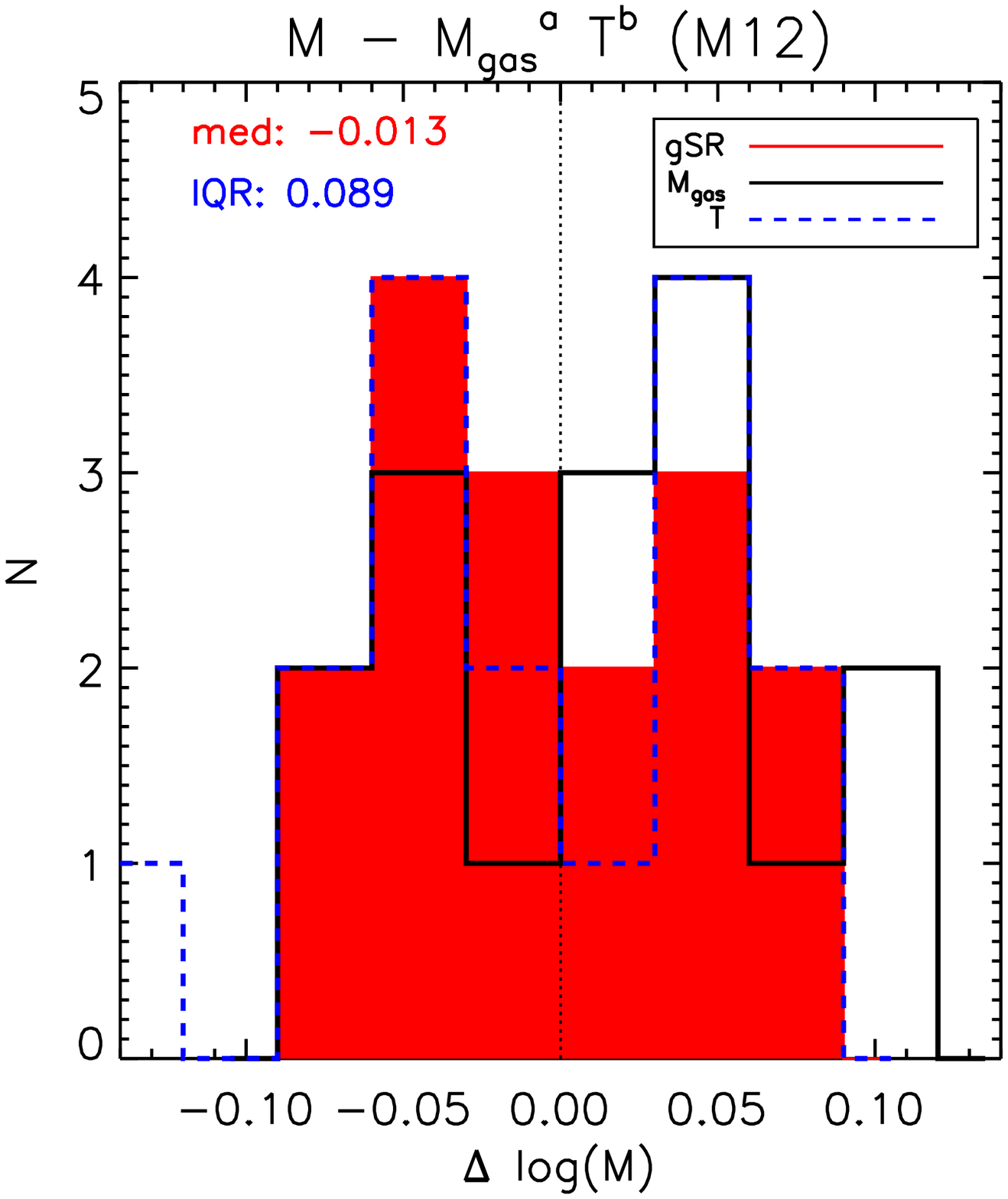}
} \caption{Best-fit results from the M12 dataset. 
} \label{fig:m12ab}
\end{figure*}

\begin{figure*}
\hbox{
 \includegraphics[width=0.33\textwidth, keepaspectratio]{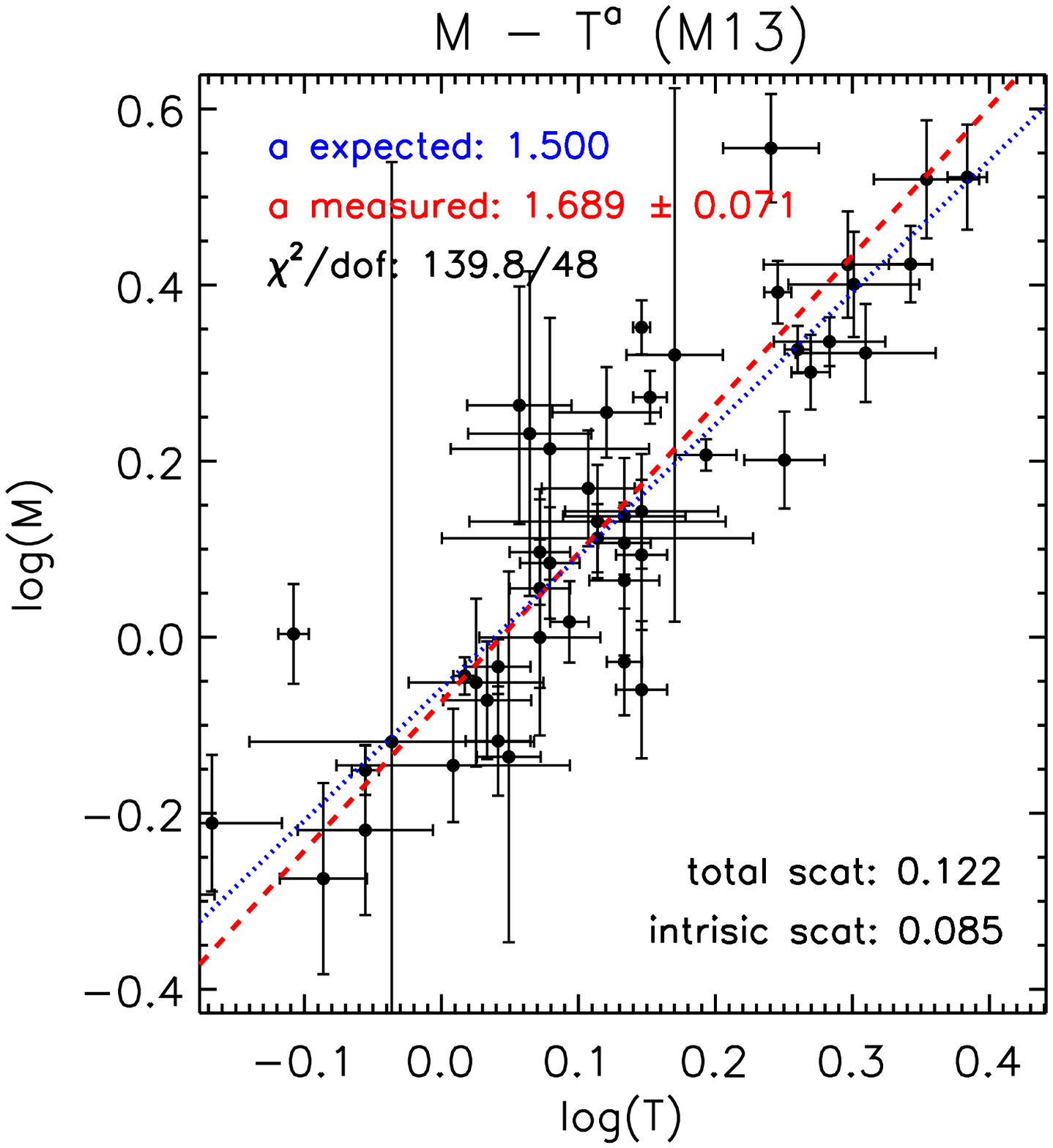}
   \includegraphics[width=0.33\textwidth, keepaspectratio]{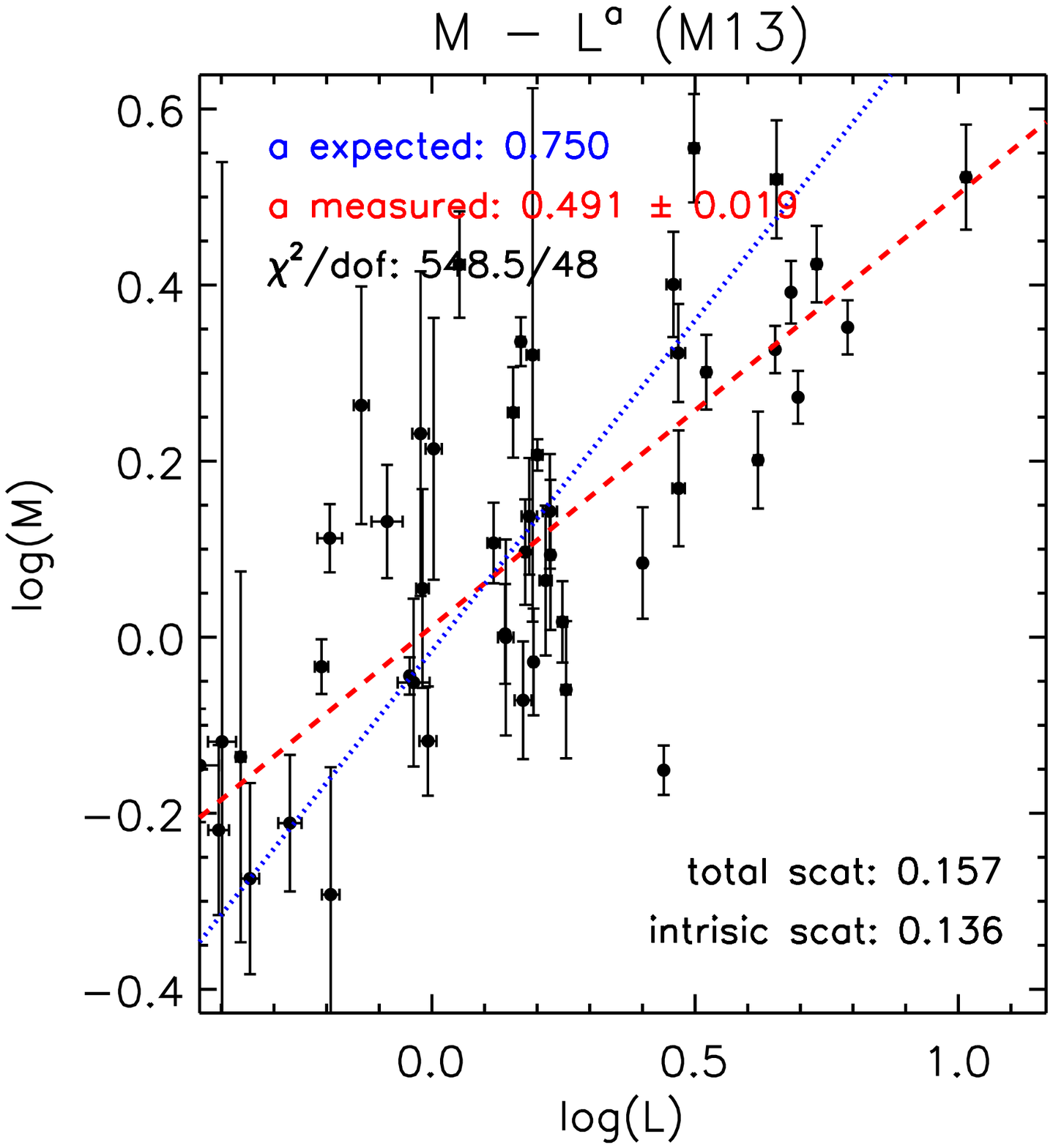}
  \includegraphics[width=0.33\textwidth, keepaspectratio]{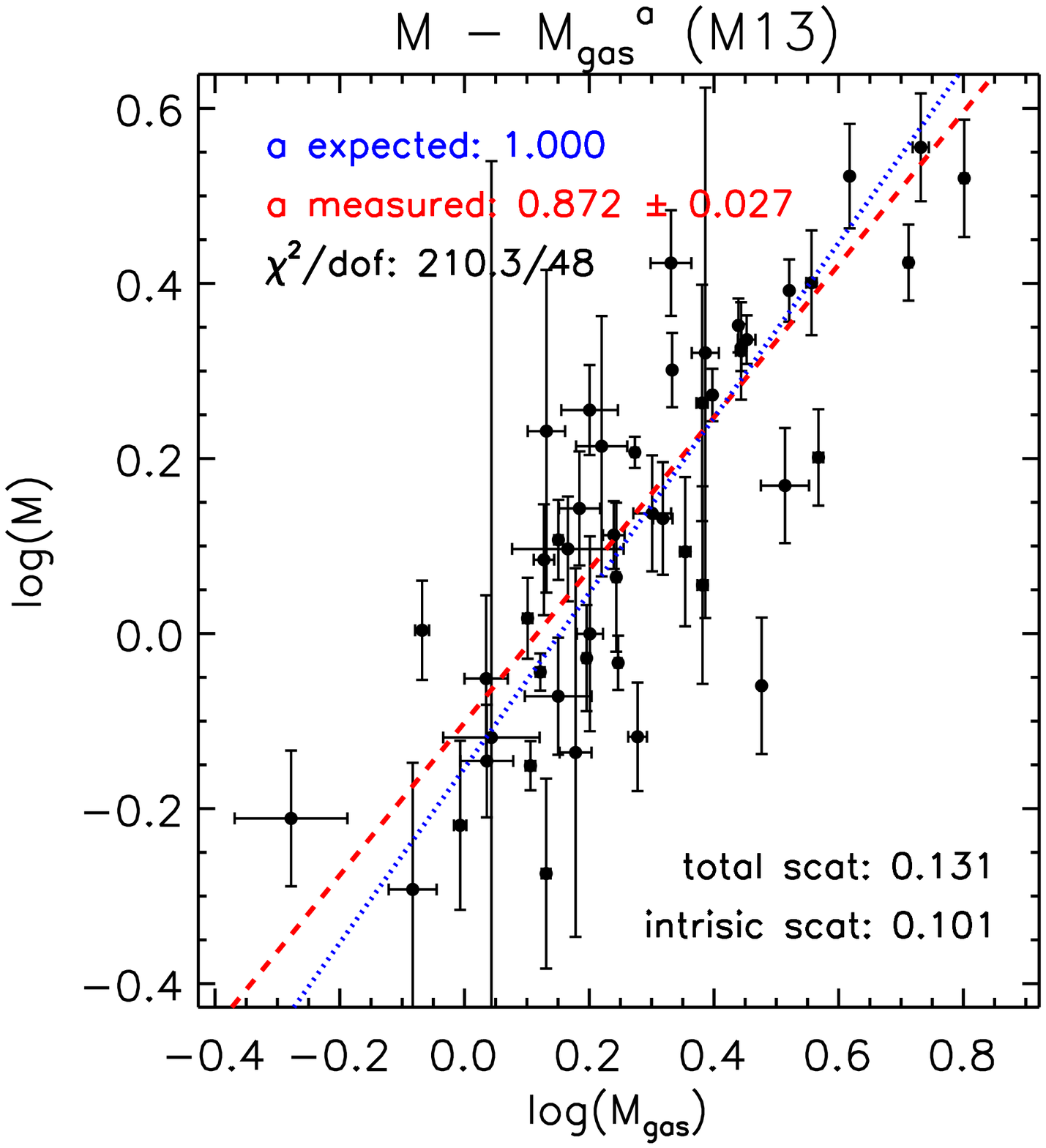}
} \hbox{
 \includegraphics[width=0.33\textwidth, keepaspectratio]{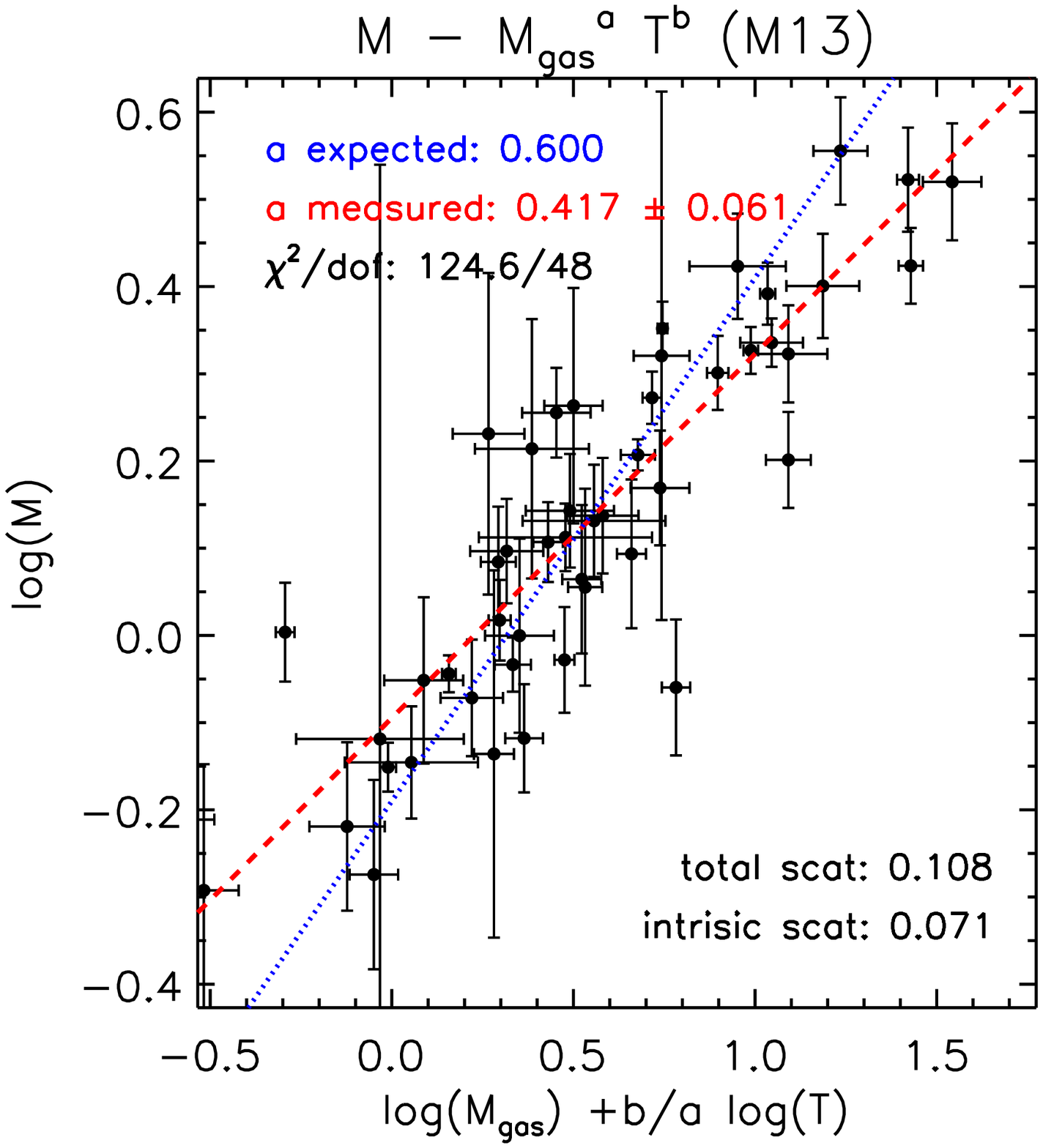}
 \includegraphics[width=0.33\textwidth, keepaspectratio]{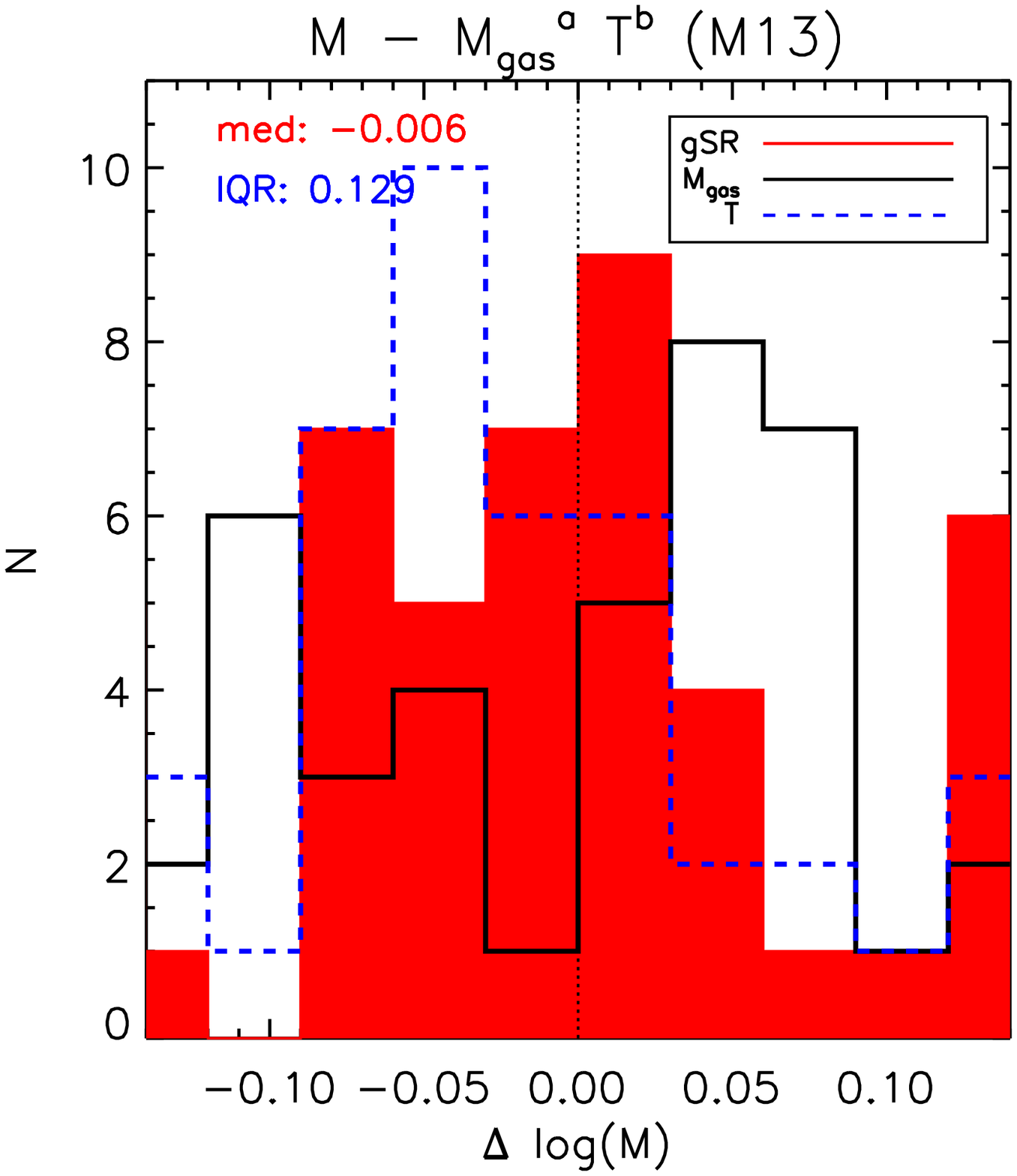}
} \hbox{ 
  \includegraphics[width=0.33\textwidth, keepaspectratio]{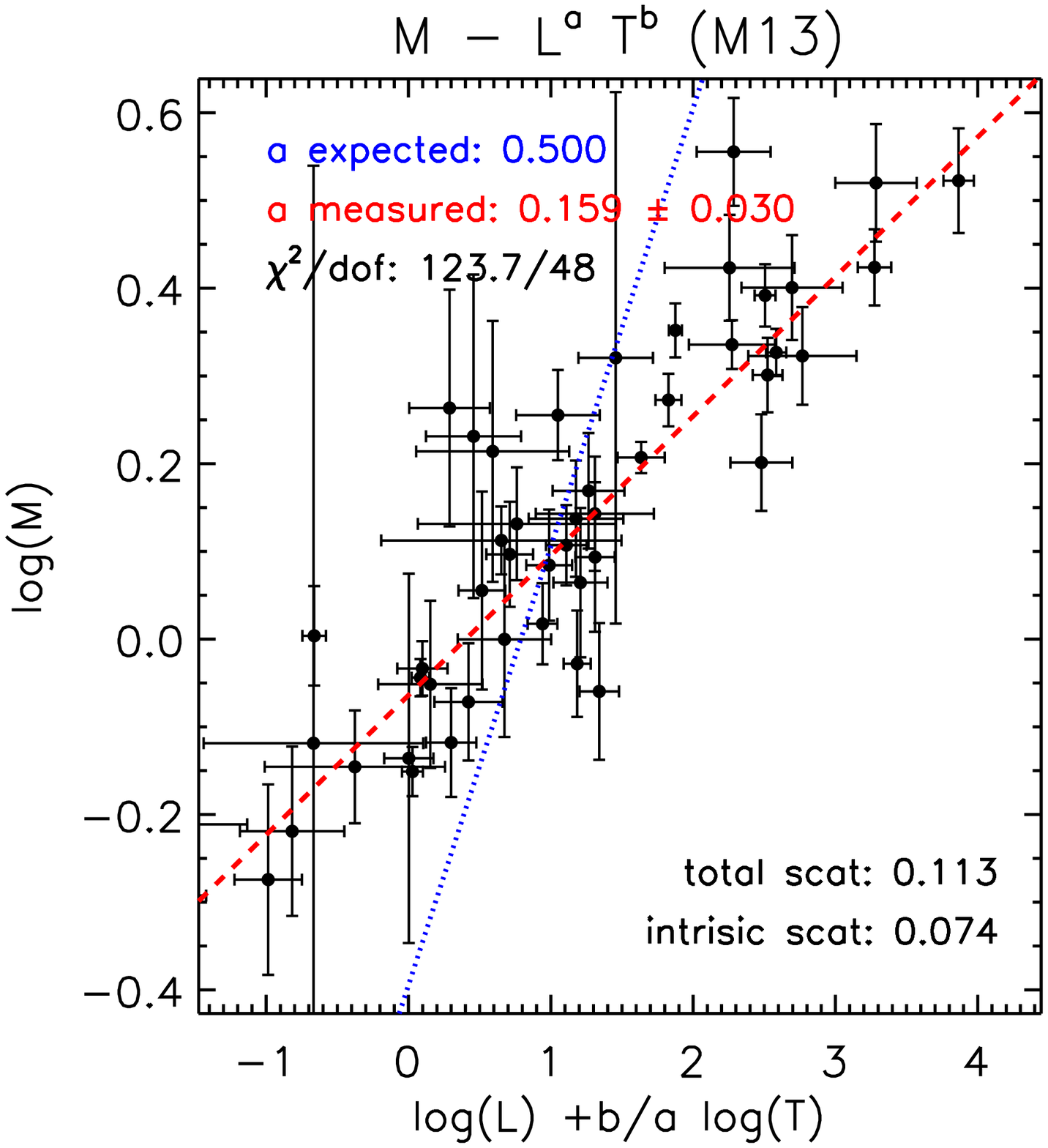}
  \includegraphics[width=0.33\textwidth, keepaspectratio]{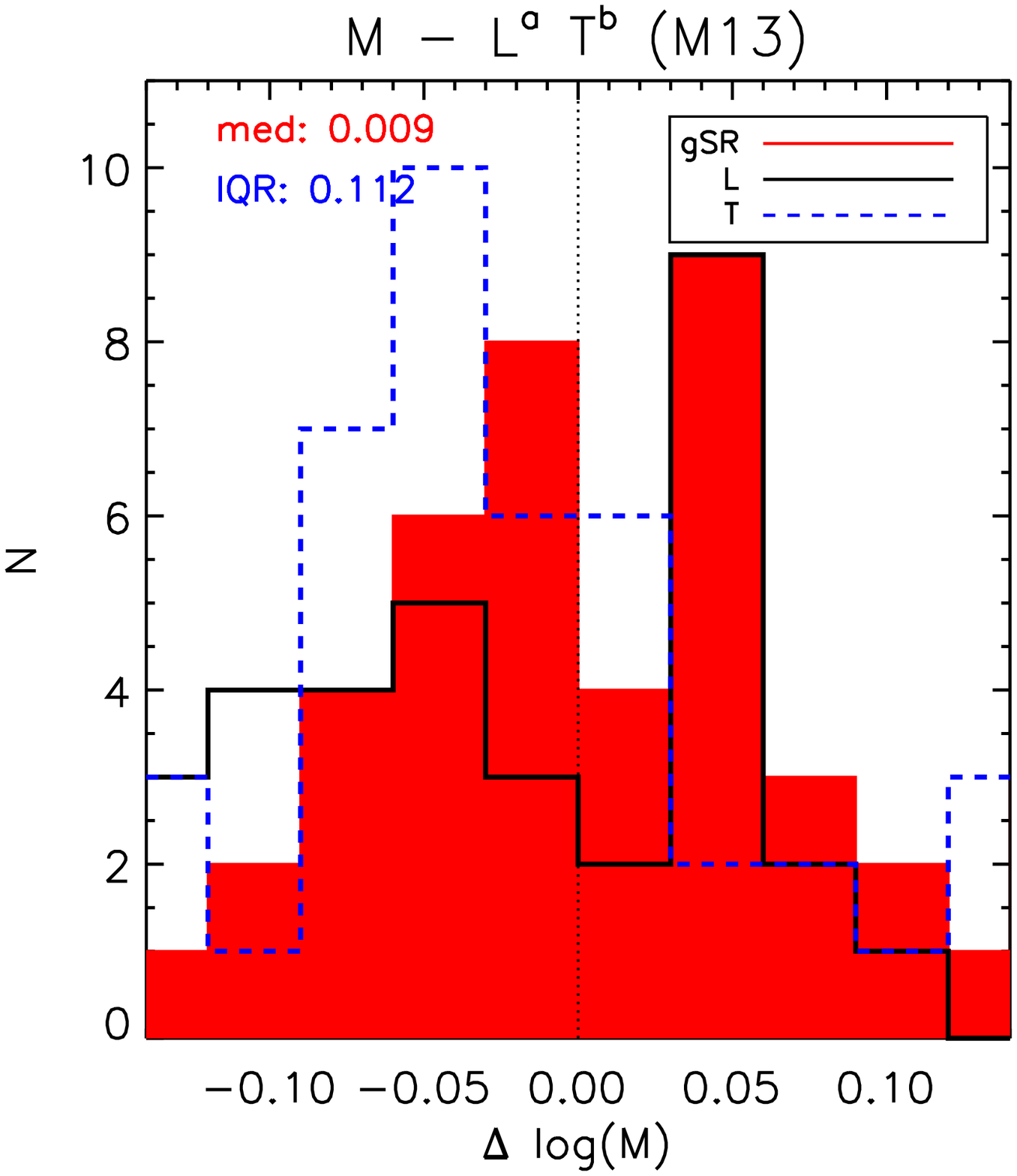}
} \caption{Best-fit results from the M13 dataset.
} \label{fig:m13ab}
\end{figure*}

\begin{figure*}
\hbox{
   \includegraphics[width=0.5\textwidth, keepaspectratio]{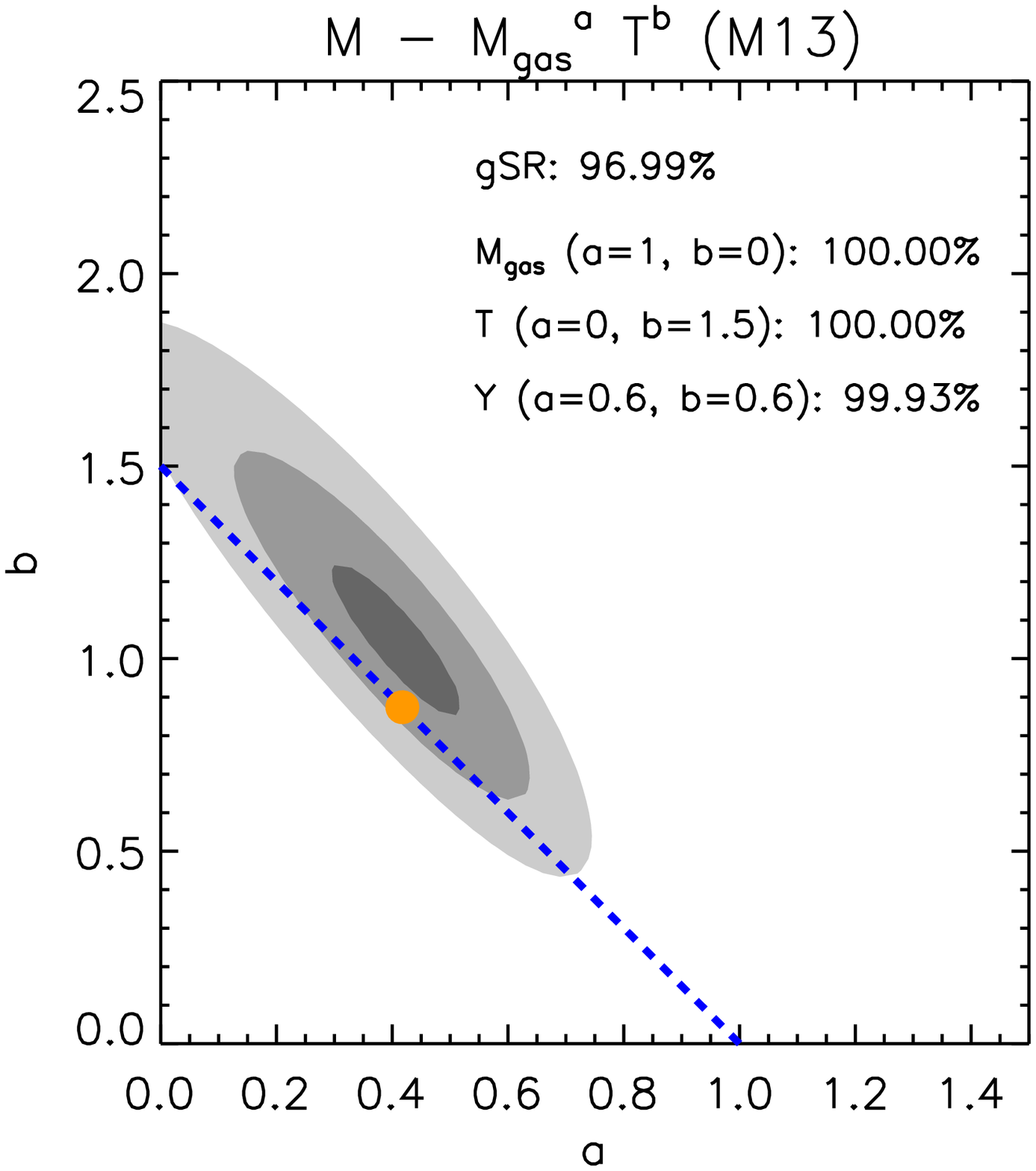}
  \includegraphics[width=0.5\textwidth, keepaspectratio]{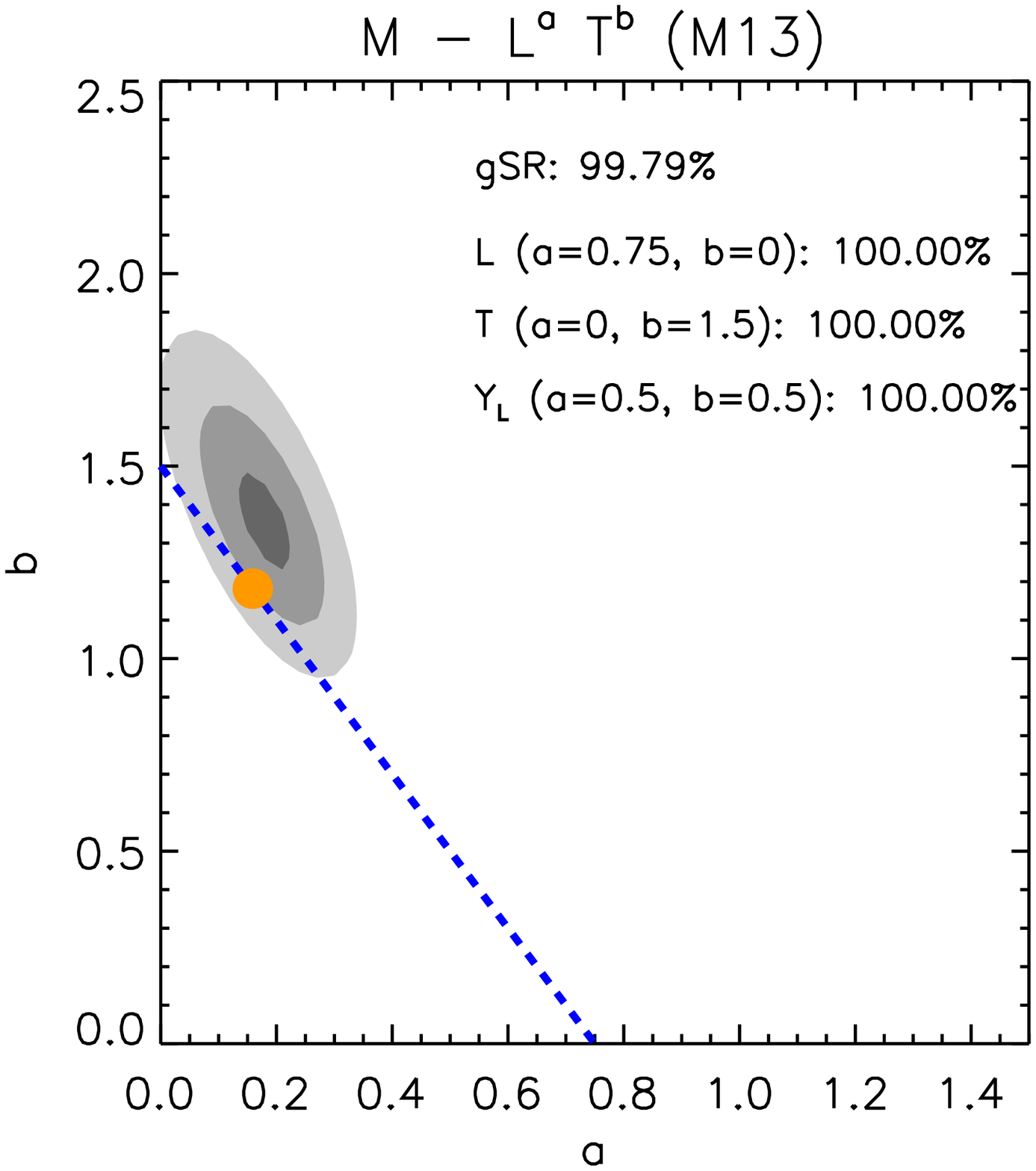}
} \hbox{
 \includegraphics[width=0.33\textwidth, keepaspectratio]{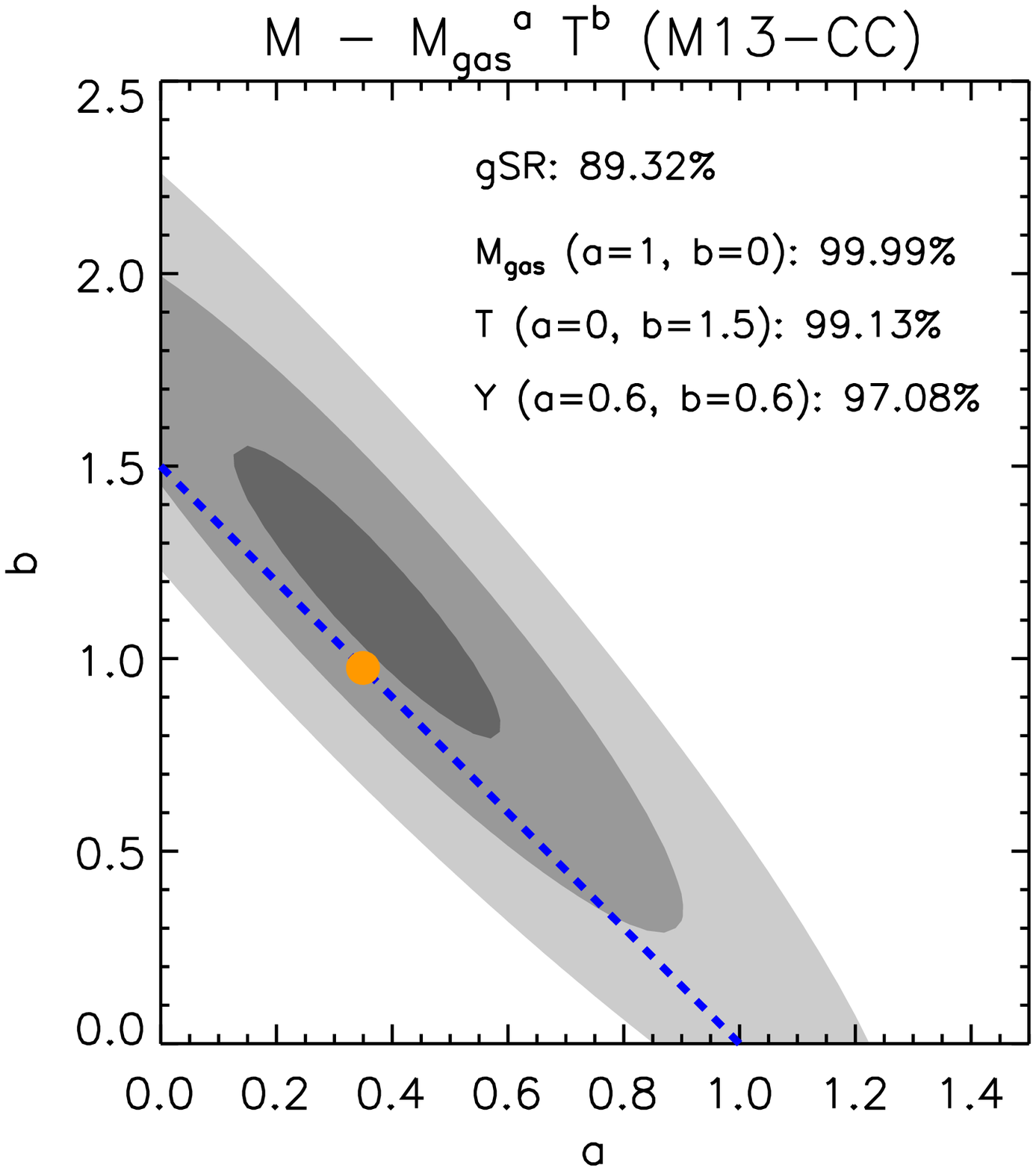}
 \includegraphics[width=0.33\textwidth, keepaspectratio]{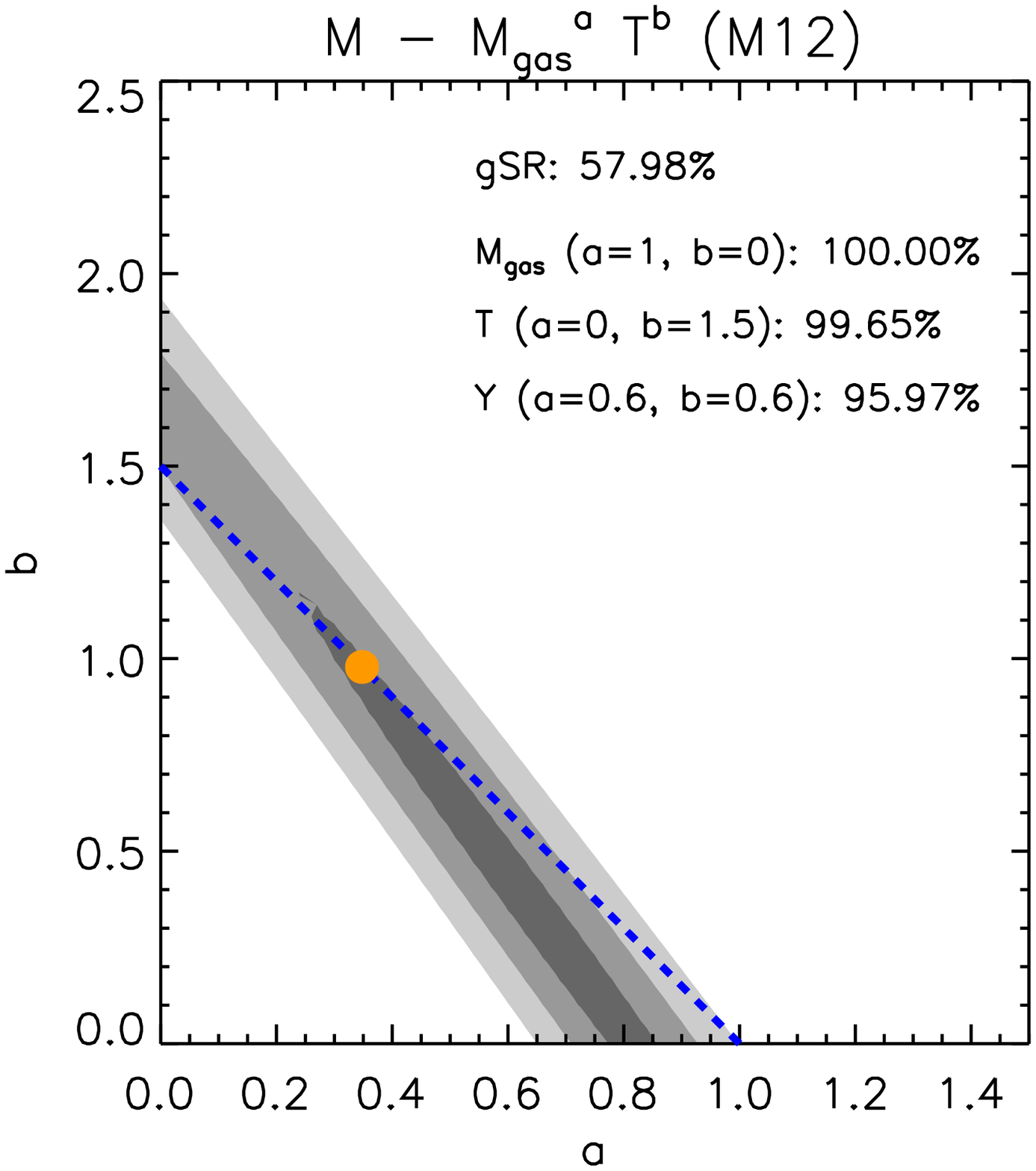}
 \includegraphics[width=0.33\textwidth, keepaspectratio]{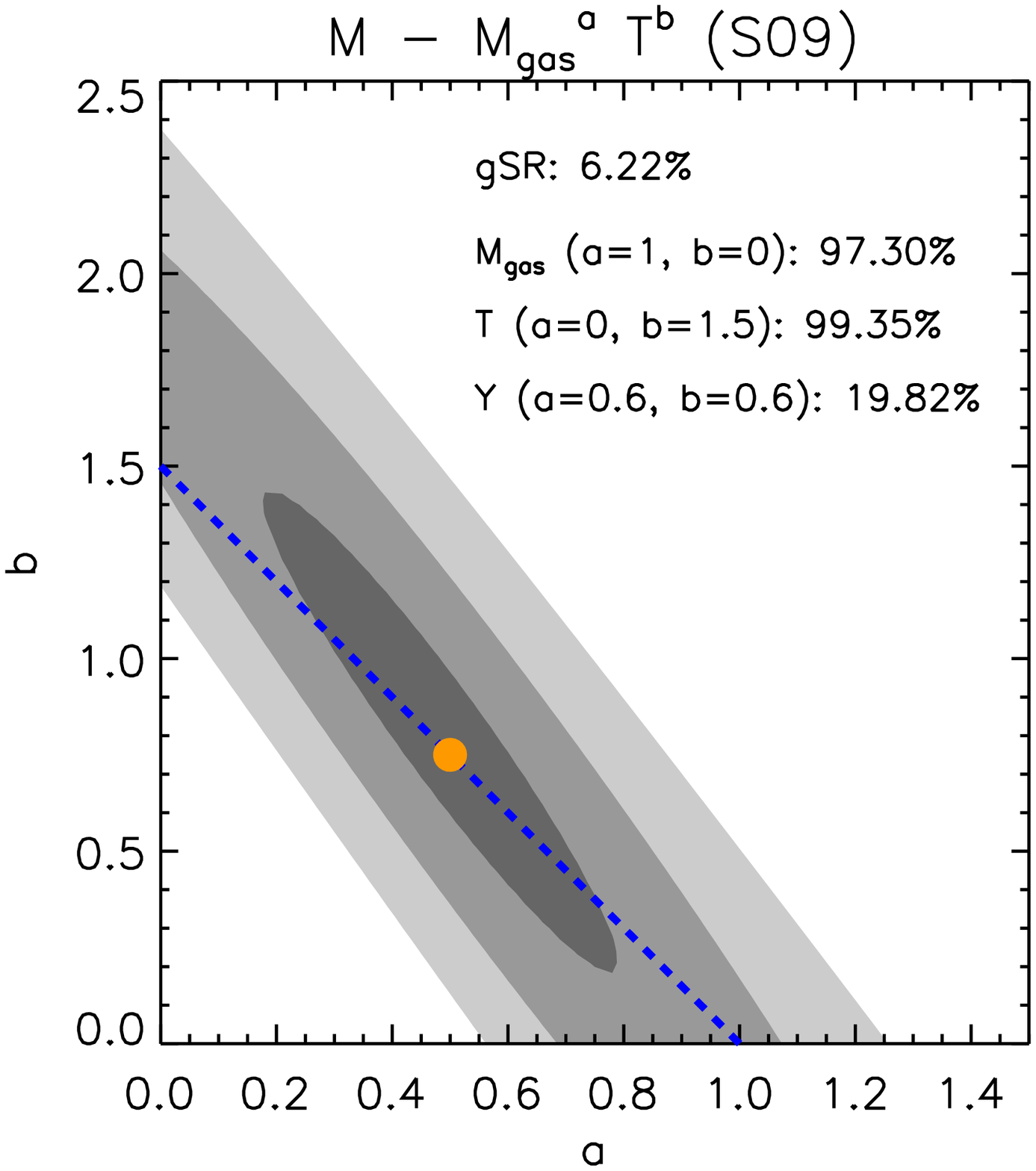}
 } 
  \caption{1 (68.3 per cent level of confidence), 3 (99.73\%) and 5 (99.9999\%) $\sigma$ likelihood contours for 2 interesting parameters ($\Delta \chi^2 = 2.3, 11.8, 28.76$, respectively). These contours are obtained by minimizing equation~\ref{eq:chi2} over a grid of $(a, b)$. By construction, the best-fit results from the gSR (orange dot) are in correspondence of the intersection between the predicted behaviour for the the self-similar prediction (blue dashed line) and the contours. Labels in the plot indicate the level of confidence (in percentage) by which the quoted solutions deviate from the minimum $\chi^2$ in the $\{a, b\}$ plane. The ``gSR'' solution refers to the result obtained by imposing the relation in equation~\ref{eq:ab} (dashed blue line). The properties of the indicated samples (M13, M13-CC, M12, S09) are listed in Table~\ref{tab:prop}.
} \label{fig:con}
\end{figure*}

\subsection{From the best-fit results to the total mass}

From the best-fit results $\{n, a\}$ obtained from the application of equations~\ref{eq:log} and \ref{eq:chi2} 
to a sample where the hydrostatic masses $M_{\rm HSE} \equiv M_{\rm tot}$ are available (see next section), it is now possible to recover
an estimate of the total gravitating mass $M_{\rm fit}$
\begin{equation}
E_z M_{\rm fit} = 10^n \; A^a \; B^b 
\label{eq:mfitab}
\end{equation}
where $b$ is related to $a$ via equation~\ref{eq:ab}.

In the present analysis, the quantities $A$ and $B$ are estimated at $R_{\Delta} = R_{500}$.
However, in general, they can also be observed at an arbitrary radius $R_0$ which is chosen, for instance, 
because encloses the region with the highest signal-to-noise ratio and is not expected to coincide 
with $R_{\Delta}$. We refer to the appendix for a discussion of the case $R_0 \neq R_{\Delta}$.

The error on $M_{fit}$ is formally due to the sum in quadrature of the propagated uncertainty obtained from the best-fit
parameters and the statistical error associated with the observed quantities.
Hereafter, we only consider the former, that can be in some way considered as a {\it systematic} uncertainty related to the set of data used to calibrate the generalized scaling relations. 

Given the best-fit parameters $\{n, a\}$ with a corresponding $2\times2$ covariance matrix $\Theta$ with elements $\Theta_{00} = \epsilon_n^2, \,\Theta_{11} = \epsilon_a^2, \, \Theta_{10}=\Theta_{01}=cov_{na}$, the uncertainty $\epsilon_M$ on $M_{fit}$ can be written as
\begin{align}
\epsilon_M & = \frac{M_{fit}}{E_z} \left( \theta_n^2 \; \Theta_{00} \; + \theta_a^2 \; \Theta_{11} \; + 2 \theta_n \theta_a \; \Theta_{10} \right)^{0.5}, \nonumber \\
 \theta_n & = \ln(10) \nonumber \\
 \theta_a & =  \ln\left( A \; B^{-(1+0.5d)} \right),
 \label{eq:errm}
\end{align}
where $\theta_n$ and $\theta_a$ indicate the partial derivative of $M_{fit}$ with respect to the best-fit parameters.
We note that the third addendum in the definition of $\epsilon_M$, which includes the off-diagonal element $cov_{na}$, is comparable in magnitude to the other two contributions and thus cannot be neglected in the total error budget measurement.

\section{The calibration of the scaling relations}
 
To calibrate the (generalized) scaling laws, we decide to analyze in a homogenous and reproducible way some published dataset. 
We search in the literature for samples with measured set of X-ray determined total mass $M_{\rm HSE}$, temperature $T$ and either gas mass $M_g$ or bolometric luminosity $L$. 

In the present work, we consider X-ray mass estimates obtained though the application of the equation of the hydrostatic equilibrium under the assumptions that any gas velocity is zero and that the ICM is distributed in a spherically-symmetric way into the cluster gravitational potential (see, e.g., Ettori et al. 2013). Considering that these conditions are verified more strictly in dynamically relaxed objects, we also use, when available, the information on the dynamical state of the objects, considering, for instance, if they are relaxed or with a cooling core (Cool Core --CC-- objects are galaxy clusters where the X-ray core has an estimated cooling time lower than the age of the structure; in general, these systems present a X-ray surface brightness map with a round shape and with no evidence of significant subclumps).
For our purpose, "CC clusters" and "relaxed clusters" identify the same category of objects for which the hydrostatic masses are more reliable.
As a result of our analysis, we discuss also any deviation in the mass reconstruction of CC/relaxed and NCC/disturbed clusters.

\subsection{The X-ray cluster samples}
\label{sect:samples}

We have selected samples over a wide range of masses to calibrate the gSR on group and cluster mass scales.
Moreover, we have considered samples in which the extrapolation over the radial range of the observed profiles of gas density and temperature has been minimal to recover the mass at $R_{500}$.
The following samples, with the main properties listed in Table~\ref{tab:prop}, are then considered:

\begin{itemize}

\item Mahdavi et al. (2013; hereafter M13): it is a compilation of 50 rich galaxy clusters with X-ray properties ($M_g$, global bolometric $L$, global emission-weighted $T$ and X-ray masses from the hydrostatic equilibrium equation $M_{\rm HSE}$) measured with \cxo\ and \xmm. All the quantities are estimated at $R_{500}$ as evaluated from the weak-lensing mass measurements, including also the core emission.
Measures of substructure help to quantify the level of departure from equilibrium and of the bias associated to the hydrostatic mass reconstruction. 
Mahdavi et al. (2013; see sections~3 and 7) found a significant correlation among all the considered substructures estimators (central entropy $K_0$, Brightest Central Galaxy to X-ray peak offset, centroid shift variance, power ratios; see also B\"ohringer et al. 2010 and Cassano et al. 2010) and concluded that the central entropy and the BCG to X-ray peak offset provide to most stringent evidence for bimodality in the cluster population between CC/relaxed and NCC/disturbed objects.
Following this result, we have also considered the two complementary sub-samples of the 16 cool core systems, identified from their quoted level of the central gas entropy (entropy value at 20 kpc $K_0 < 70$ keV cm$^2$), and of the remaining 34 objects. These two sub-samples were labelled M13-CC and M13-NCC, respectively.

\item Maughan (2012; M12): this is a compilation of 16 nearby massive objects with measured $M_g$, gas temperature $T$ in the $[0.15 - 1] R_{500}$ aperture and $M_{\rm HSE}$ at $\Delta=500$.
These objects were selected from the samples described in Vikhlinin et al. (2006) and Arnaud et al. (2007) for their precise mass estimates from X-ray hydrostatic analyses.
Vikhlinin et al. (2006) present the mass profiles, derived from \cxo\ exposures, for 13 low-redshift, relaxed clusters (with the only possible exception of A2390 that has an ICM emission not spherically symmetric nor expected to be in hydrostatic equilibrium) in a temperature interval of 0.7--9 keV. All the gas density and temperature profiles of the nine clusters considered in the M12 sample extend almost to $R_{500}$, permitting a mass estimate without any extrapolation.
Arnaud et al. (2007) discuss the mass profiles in ten nearby morphologically relaxed clusters over the temperature range 2--9 keV and observed with \xmm. The quoted $M_{500}$ of the seven clusters considered in the M12 sample were derived from the mass profiles measured to overdensities of about 600-700, apart from the two coolest systems (at overdensity of $\sim$ 1400).
All the objects in the M12 sample are relaxed systems and are thus labelled as ``CC'' objects (see Table~\ref{tab:mfit}).

\item Pratt et al. (2009; P09): this work quotes gas mass $M_g$, bolometric luminosity and spectroscopic temperature both within $R_{500}$ and in the $[0.15-1] \; R_{500}$ region for the 31 nearby clusters part of the Representative \xmm\ Cluster Structure Survey (REXCESS). 
The estimates of $R_{500}$ are obtained from the best-fit constraints of the $M-Y_X$ relation in Arnaud et al. (2007). 
Note that the $M_{\rm tot}$ considered for this sample are not direct measures of the hydrostatic mass but are obtained from the quoted estimates of $R_{500}$.
We use them just for comparison and not to calibrate the gSR. In the following analysis, we refer to $L$ and $T$ as the values estimated within $R_{500}$. 
Out of 31 objects, twelve were classified as morphologically disturbed because have a centroid shift, that measures the standard deviation of the projected separation between the X-ray peak and the X-ray centroid, larger than $0.01 R_{500}$. 

\item Sun et al. (2009; S09) present a systematic analysis of 43 nearby galaxy groups observed with \cxo. We have considered the 23 objects for which the gas properties (specifically $M_{\rm HSE}, M_g$ and $T$) can be measured, even with a mild extrapolation, up to $R_{500}$, 
that is determined from the application of the equation of the hydrostatic equilibrium using the best-fit functional forms of the three-dimensional gas temperature and density profiles.
This sample includes the 11 objects in {\it Tier 1}, where the X-ray surface brightness is derived at $>2 \sigma$ level to $r>R_{500}$ and the gas temperature profile extends up to $r > 0.8 R_{500}$, and the 12 groups in {\it Tier 2}, with surface brightness and temperature profiles available to at least $R_{1000} \approx 0.7 R_{500}$.
The adopted gas temperatures are obtained as projection of the integral of the three-dimensional profile over the radial range $[0.15-1] \; R_{500}$.  
Because this sample has been selected to have the X-ray emission centered around the central galaxy and not significantly elongated nor disturbed beyond the group core, we qualify all of them as CC/relaxed systems.

\end{itemize}

All the physical quantities considered here refer to the cosmological parameters $H_0=70$ km s$^{-1}$ Mpc$^{-1}$ and $\Omega_{\rm m} = 1 - \Omega_{\Lambda}=0.3$. For only one sample (S09), a conversion from an other cosmological framework has been required.
In this case, we use the relations $M_{\rm HSE} \propto d_{\rm ang}$ and $M_g \propto d_{\rm ang}^{2.5}$, where $d_{\rm ang}$ is the angular diameter distance, to make the proper 
conversion.
As described above, the radius of reference for the present analysis is $R_{500}$. Note that the samples here considered use different techniques to measure it: M12 and S09 recover $R_{500}$ from the hydrostatic mass profile; M13 uses the result from the weak-lensing analysis; P09 applies the $M-Y_X$ scaling relation.
Considering that we will analyze each sample independently, the use of different definitions of $R_{500}$ will permit us to test further the performance of the gSR.

Note also that 6 objects (MKW4, Abell2717, Abell1991, Abell2204, Abell383, Abell2390) are in common to different samples. The quoted hydrostatic masses show differences  between $0.2 \sigma$ and $\sim 2 \sigma$, with the most deviant values for MKW4 and Abell383. For MKW4, the difference between the hydrostatic masses in S09 and M12 (as adopted from Vikhlinin et al. 2006), where the same \cxo\ dataset is used, is discussed in the Appendix of Sun et al. (2009) and is probably due to a different modelling of the gas density profile. In the case of Abell383, the difference between the values quoted in Mahdavi et al. (2013), which is based on a joint analysis of the \xmm/\cxo\ exposures, and Vikhlinin et al. (2006), which analyze only the \cxo\ data, can be explained, at least partially, with both the different dataset used and the different estimate of $R_{500}$ where the total mass is evaluated. Indeed, in M13, $R_{500}$  is adopted from the result of the weak-lensing analysis and is about 7 per cent larger than in M12, implying $M_{500}$ higher by $\ga$ 20 percent.

\subsection{The best-fit results}

To compare the performance of the gSR versus the standard relations, we focus our study on the following relations: $M_{\rm HSE}-T$, $M_{\rm HSE}-M_g$, $M_{\rm HSE}-L$ and the gSR $M_{\rm HSE} - M_g T$, $M_{\rm HSE} - L T$.
As an example, we show in Fig.~\ref{fig:m12ab} and \ref{fig:m13ab} the best-fit lines and the distribution of the residuals for the M12 and M13 sample, respectively.
The distribution of the residuals in $\log(M)$ shows an appreciable reduction of both the median deviation and the Inter-Quartile-Range for the clusters in, e.g., M13. 
No clear improvements are noticed for M12, where the measured intrinsic scatter is already close to zero when the standard scaling laws are applied.

In Fig.~\ref{fig:con}, we plot the likelihood contours obtained for a grid of values of the slopes $\{a, b\}$. 
These statistical constraints show the locus of the slopes preferred from the data in terms of the minimal $\chi^2$. 
This locus can be well approximated by the relation identified in the hydrodynamical simulations discussed in E12 and indicated by equation~\ref{eq:ab} (and equation~\ref{eq:proj}).
In the same figure, we also show the significance of the deviation from the minimum value of the $\chi^2$ for the most interesting cases, nominally the best-fit values obtained by imposing equation~\ref{eq:ab} and the standard self-similar relations.
We notice how the latter relations that make use of either the gas temperature or the gas mass only are systematically above the lowest value of $\chi^2$ at a level of confidence $>99$ per cent. 
Only the cases where the $Y_X = M_g \; T$ quantity is adopted provide less significant deviations, but always in the order of 95 per cent (about $2 \sigma$ for a Gaussian distribution) or larger. The only exception is the sample S09, where the total mass can be recovered using $Y_X$ at a level of confidence of $\sim 20$ per cent. 
However, the gSR provides always the best performance, with the significance of the deviations from the absolute minimum in the $\{a, b\}$ plane ranging from only 6 per cent (S09 sample) to 99.8 per cent (M13 sample using $LT$).

We present the best-fit results in Tables~\ref{tab:bestfit_scat}, \ref{tab:bestfit_par} and \ref{tab:bestfit_ratio}.
For sake of completeness, we include also the case $M - L M_g$ (see equation~\ref{eq:proj}), showing how this relation provides a scatter in reconstructing the total mass higher than the two other relations investigated ($M-M_g T$ and $M-L T$) and, therefore, will be not discussed further.

When the standard scaling laws are used,
the M13 sample shows the lowest $\chi^2$ for the $M-T$ relation, whereas M12 and S09 seem to prefer slightly the $M-M_g$ one.
When a gSR is applied, we measure systematically a reduction of the total $\chi^2$, with improvements in $\Delta \chi^2$ up to 16--325 (with 48 dof) in M13, for all the datasets analyzed here.
Even in the case where a significant reduction in $\chi^2$ is not observed (as for the M12 sample), we measure a reduction of the total scatter of $\ga 20$ per cent.
The intrinsic scatter associated to the best-fit with a gSR is below 0.07 (corresponding to a relative error on the total mass lower than 16\%) in all cases.

In general, we measure a reduction in the $\chi^2$ value, total and intrinsic scatter when a gSR is adopted in place of a standard self-similar relation (see Table~\ref{tab:bestfit_scat}). 
We use this evidence as confirmation that gSR reproduces better the distribution of the estimated hydrostatic mass used to calibrate these relations.
We note also that the M13 sample presents the largest total and intrinsic scatter among the analyzed datasets. 
This might be related also to the use of $R_{500}$ as obtained from the weak-lensing analysis, whereas a X-ray based definition of $R_{500}$, and therefore correlated to the quantities investigated, is adopted for the other samples.

The best-fit results (Table~ \ref{tab:bestfit_par}) for the samples M13, M12 and S09 agree on the slope $a$ of the $M_{\rm HSE} - M_g T$ relation (the error-weighted mean is $a = 0.411 \pm 0.050$). The slope of the $M_{\rm HSE} - L T$ relation is about $0.15$. 
On the other hand, we notice significant differences in the normalization of the $M_g T$ gSR between M12 ($n = -0.054$) and M13 ($n = -0.095$) that induce estimates of masses larger by about 10 per cent when the best-fit results from M12 are adopted. 
This can be explained by the fact that the objects in M12 are all relaxed systems, whereas the M13 sample is more heterogeneous (see also discussion in Mahdavi et al. 2013), including both relaxed and dynamically disturbed systems.  
The hydrostatic mass in the latter ones is indeed expected to underestimate the true mass due to an uncounted contribution from residual bulk motions of the ICM to the total energy budget (e.g. Nelson et al. 2012, Rasia et al. 2012, Suto et al. 2013). If we consider only the sub-sample of relaxed objects in M13-CC, we measure a normalization of the $M_{\rm HSE} - M_g T$ relation that matches (within $1 \sigma$) the value measured for the M12 dataset. 

We further confirm this evidence by quantifying it in Table~\ref{tab:bestfit_ratio}, where we present the ratios between the estimates of the mass recovered from the best-ft gSR and the input hydrostatic masses. All the deviations are in the order of few per cent when the $M_{\rm HSE}$ are recovered within the same sample and, on average, of about 10\%, with a dispersion of $\sim$20\%, when different sample are used.

In particular, the gSR defined with $M_g$ and $T$ and calibrated with M13 reproduces the mass estimates in M12, P09 and S09 with $(M_{\rm fit} - M_{\rm HSE}) / M_{\rm HSE} = \Delta M / M_{\rm HSE}$ of --10, --4 and --7 per cent, respectively. When the gSR is calibrated with the M12 sample, the mass measurements in M13 are recovered with $\Delta M/ M_{\rm HSE} \sim +$12 per cent and the ones in S09 and P09 with a mean ratio of $+$6 per cent. Using the M13-CC sample provides similar results, with deviations in the order of $+$10 per cent for the data in M13 and of few per cent the masses quoted in M12, S09 and P09.
The same sub-sample induces over-estimates of the hydrostatic mass in disturbed objects (collected in the sub-sample M13-NCC) by 19 per cent on average, as produced from the $M_g \, T$ gSR calibrated with M12.

When the $L T$ gSR calibrated with either M13 or M13-CC is used, we measure deviations lower than 5 per cent in the reconstructed hydrostatic mass $M_{fit}$ of the clusters in the M13, M13-CC and,  curiously, even in the M13-NCC sample.
This result, which shows that the original hydrostatic masses in disturbed objects are well recovered, on average, when $L T$ gSR is calibrated with samples containing CC systems, appears at odd with the previous evidence that $M_g T$ gSR provides values of the $M_{\rm fit}$ of NCC clusters that are higher than their $M_{\rm HSE}$.
To explain this, we have to consider that NCC clusters present a higher entropy level in the core with respect to the more relaxed systems (e.g. Mahdavi et al. 2013), due to the phenomena (such as merging events) that disturb their X-ray emitting plasma. As consequence of that, the global gas luminosity, in particular when the core is not excised as in M13, is lower, for a given mass halo, than the one measured in CC clusters used to calibrate the $L T$ gSR. Hence, this relation will provide a $M_{\rm fit}$ lower for a NCC than for a CC, almost compensating for the above-mentioned bias on the hydrostatic mass and matching the tabulated $M_{\rm HSE}$.

Deviations of few per cent are also measured when the masses in P09 are reconstructed. 
If we consider for this sample $L$ and $T$ extracted over the region $[0.15-1] R_{500}$ (i.e. excluding the core emission), 
we obtain larger deviations (in the order of $-$12 and $-$7 per cent, as mean values, using calibration provided from M13 and M13-CC, respectively), 
because the luminosities considered for the M13 sample are not core-excised and, therefore, are higher at a given mass. 

On the galaxy group scales, using the S09 sample to calibrate the $M_g T$ gSR, we measure deviations, on average, between 0 and 7 per cent for the samples M13-CC, M12 and P09, with larger values of $\sim +$15 per cent for M13. These values indicate that the gSR, although tuned to systems with mean total mass about 6--8 times lower than the ones in M12 and M13, is able to reproduce the measured $M_{\rm HSE}$ in these samples, showing a bias that is due to the fact that the S09 sample is dominated by relaxed systems.

\section{Summary and discussion}

In this work, we have discussed the application of the generalized scaling relations presented in Ettori et al. (2012) to real data.
In the context of the self-similar model for X-ray galaxy clusters, we show that a generic relation between the total mass and a set of observables like gas luminosity, mass and temperature can be written as $M_{\rm tot} \propto L^{\alpha} M_g^{\beta} T^{\gamma}$, where the values of the slopes satisfy the relation $4 \alpha +3 \beta +2 \gamma = 3$ (and $M_{\rm tot} \equiv M_{\rm HSE}$ by the definition adopted in the present work). 
Some projections of this plane are particularly useful in looking for a minimum scatter between X-ray observables and hydrostatic mass:
$M_{\rm tot} \propto A^a B^b$, where $A$ is either $M_g$ or $L$, $B = T$ and $b = 1.5 \; -(1 \;+0.5 d) \; a$, with $d$ equals to the power to which the gas density appears in the formula of the gas mass ($d = 1$) and luminosity ($d = 2$).

We show indeed that the gSR are the most efficient relations, holding among observed physical quantities in the X-ray band, to recover the gravitating mass on both galaxy group and cluster scales, because they provide the lower values of $\chi^2$, total and intrinsic scatter among the studied scaling laws. The intrinsic scatter associated to the best-fit with a gSR at $\Delta=500$ is below 0.07 (corresponding to a relative error on the total mass lower than 16\%) in all cases.

The best-fit results on the different samples considered in our analysis agree on the slope $a$ of the $M_g T$ gSR (the error-weighted mean is $a = 0.41 \pm 0.05$) and are consistent for the slope of the $L T$ relation (the error-weighted mean is $a \approx 0.15$). 
These values are significantly different from any adopted relations so far (e.g. $M\propto M_g$ requires $a=1$, $M \propto T^{3/2}$ needs $a=0$, $M \propto Y_X$ is obtained for $a=0.6$).
This demonstrates that, still in the self-similar scenario, the gSR provides more flexible tool to use the X-ray observables as robust X-ray mass proxies.
In particular, our best-fit results on the slope prefer a larger contribution from the gas global temperature than from the gas mass or luminosities.
However, we show that the latter ones are needed to optimize the mass calibration. 
The combination of the constraints from the depth of the halo gravitational potential
(through the gas temperature $T$) and from the distribution of the gas density
(traced by $M_g$ and the X-ray luminosity), that is more prone to the ongoing
physical processes shaping the ICM global properties, is therefore essential to 
link the cluster X-ray observables to the total mass.

Nonetheless, we notice a significant difference in the normalization of the $M_g T$ gSR between the fit obtained with data in Maughan (2012), that includes only relaxed systems, and that based on the Mahdavi et al. (2013) sample, that, on the contrary, is dominated (68 per cent) from disturbed objects. This difference induces estimates of masses larger by about 10 per cent when the best-fit results from Maughan (2012) are adopted and is reduced when the sub-sample of relaxed clusters from M13 is considered. 
Samples dominated by relaxed systems (as in M13-CC, M12, S09) provide calibrations of the $M_g T$ gSR that tend  to over-estimate the hydrostatic mass in disturbed objects (M13-NCC)  systematically by a mean value of 18--24 per cent.
Indeed, in not-relaxed clusters, a non-thermal component is expected to contribute to the total energy budget, biasing low the estimate of the X-ray mass as traced though the hydrostatic equilibrium equation (e.g. Nelson et al. 2012, Rasia et al. 2012, Suto et al. 2013). 
Thus, the results quoted above seem to confirm that, in NCC systems, the total mass as estimated through the hydrostatic equation is under-estimated, on average, by 18--24 per cent. 
The measured bias is consistent with the results discussed in Mahdavi et al. (2013) where estimates of hydrostatic and weak lensing masses are compared. They conclude that (i) these estimates are similar in CC clusters and (ii) hydrostatic masses in NCC clusters are lower by 15--20 per cent.
Using different mass proxies is definitely the most robust approach to constraint the level of mismatch on the gravitating mass between relaxed and disturbed galaxy clusters.
Several observational biases can indeed play a significant role to assess the differences in mass between relaxed and disturbed objects using X-ray scaling relations only.
For instance, it has been recognized that hydrostatic bias is composed from two main components, one related to the non-thermal source of extra-pressure and the other to temperature inhomogeneities in the ICM (see, e.g., discussion in Rasia et al. 2012). 
The acceleration of the gas becomes also a non-negligible component of the hydrostatic bias in the cluster outskirts (Suto et al. 2013, Lau et al. 2013).
Moreover, during the different phases of a merger, the values of the integrated physical properties, like $T$ and $L$, oscillate (see, e.g., Rowley et al. 2004, Poole et al. 2007). 
Only when a solid and confident knowledge is reached on the relative average variations in $T$, $M_g$ and $L$ at a fixed halo mass between a CC and a NCC galaxy cluster (as classified accordingly to its observational X-ray properties), the CC--calibrated gSR can be then used to evaluate a ``correct'' mass for a NCC system, where the term ``correct'' indicates the value of the hydrostatic mass once a proper thermalization of the ICM occurs.

On the contrary, $L T$ gSR calibrated with M13 and M13-CC over-predicts the masses in NCC objects only by few per cent.
In this case, we have to consider that the gas luminosity (as estimated over the whole cluster volume, i.e. not excluding any core emission) of NCC clusters tend to be lower than the one of relaxed objects that have been used to calibrate the gSR. 
This lower luminosity is the product of the higher central entropy induced from, e.g., recent mergers in disturbed, NCC systems 
(e.g. Rowley et al. 2004, Poole et al. 2007). For instance, by reducing the global bolometric $L$ by a factor of 2, and considering the slope of 0.15 that appears in the gSR, a compensation of about 10 per cent is provided to the above-mentioned hydrostatic bias, permitting to recover the estimated hydrostatic mass $M_{\rm HSE}$ for NCC clusters within a few per cent.

Moreover, when we calibrate the gSR with galaxy groups having a mean total mass about 6--8 times lower than the most massive systems studied here, we are still able to reproduce the measured $M_{\rm HSE}$ on cluster scales. A residual bias is present and due to the fact that the S09 sample used for the calibration in the present study is dominated by relaxed systems.

These generalized scaling relations can be easily applied to present (e.g. {\it XXL}, Pierre et al. 2011) and future (e.g. {\it eROSITA}, Merloni et al. 2012) surveys of X-ray galaxy clusters. Either the calibrations presented here are adopted and used to infer hydrostatic masses for a sub-set of systems with measured gas temperature and gas mass or luminosity, or new calibrations are estimated as described in this work for a subsample of objects selected to be representative of the population of the observed clusters.

As a by product of this study, we provide in Table~\ref{tab:mfit} the estimates of the gravitating mass at $\Delta=500$ for 120 objects (50 from the Mahdavi et al. 2013 sample, 16 from Maughan 2012; 31 from Pratt et al. 2009; 23 from Sun et al. 2009), 114 of which are unique entries. 
If we do not consider any uncertainty associated with the observed quantities, the typical relative error on the mass provided 
from, e.g., the $M_g T$ gSR with the considered datasets (see Table~\ref{tab:mfit}) 
ranges between 3.0 $\pm$ 1.6 per cent in M13 (with a relative uncertainty related to the residual intrinsic scatter of about 0.07 in log space, which corresponds to $16^{+4}_{-2}$  per cent on the quantity $\epsilon_M / M$) and 9.6 $\pm$ 4.1 per cent (with a null intrinsic scatter) in S09. The other samples provide typical errors in of $\sim$5 per cent (M13-CC: 6 $\pm$ 4 per cent and $14^{+6}_{-4}$ per cent from the intrinsic scatter; M12: 5 $\pm$ 2 and almost nil contribution from the intrinsic scatter).

This catalog of X-ray cluster masses can be used fruitfully, for instance, to compare results obtained with other techniques (like, e.g. lensing, galaxy velocity dispersion, caustics) or to apply statistics that want to address the presence, and the significance, of objects with extreme values in mass (e.g. Waizmann et al. 2013).

\section*{ACKNOWLEDGEMENTS} 
We thank the anonymous referee for helpful comments that improved the presentation of the work.  
We acknowledge the financial contribution from contracts ASI-INAF I/023/05/0 and I/088/06/0.  
We thank Elena Rasia and Gianni Zamorani for useful discussions.

\appendix

\section{The radial dependence of the observed quantities}

The estimate of the best-fit mass in equation~\ref{eq:mfitab} assumes that
the quantities $A$ and $B$ are observed at the radius $R_0 = R_{\Delta} \equiv R_{500}$.

In the case that $R_0 \neq R_{\Delta}$, one solution is to re-iterare the process till a convergence between $R_0$ and $R_{\Delta}$ is reached
within a given tolerance. However, this is computationally expensive and can be easily avoided by modelling
the radial dependence of the quantities of interest.
If we consider a radial correction in the form of a power-law 
\begin{align}
A = & A_0 \;  r^{\gamma} \nonumber \\
B = & B_0 \; r^{\tau},
\end{align}
and using the definition of the mass associated with an overdensity $\Delta$ within a sphere with radius $R_{\Delta}$, $M_{fit} = 4/3 \pi \rho_{c, z} \Delta R_{\Delta}^3$, we can write
\begin{align}
\widehat{\Delta} R_{\Delta}^3 & = 10^n A^a B^b \left( \frac{R_{\Delta}}{R_0} \right)^{\epsilon}   \nonumber  \\
\epsilon & = a \gamma + b \tau,
\end{align}
where $\widehat{\Delta} = \frac{4}{3} \pi \rho_{c, z} E_z \Delta$.

Finally, by inverting this expression to isolate the quantity of interest $R_{\Delta}$, we obtain the relation
\begin{equation}
R_{\Delta} = \left( \widehat{\Delta}^{-1} 10^n A^a B^b R_0^{-\epsilon} \right)^{ 1/(3-\epsilon) } 
\label{eq:rdelta}
\end{equation}
The estimated mass will be then obtained by substituting equation~\ref{eq:rdelta} in the definition of $M_{fit}$ and using equation~\ref{eq:ab}.

The error on $M_{fit}$ is formally due to the sum in quadrature of the propagated uncertainty obtained from the best-fit
parameters and the statistical error associated with the observed quantities.
Hereafter, we only consider the former, that can be in some way considered as a {\it systematic} uncertainty related to the set of data used to calibrate the generalized scaling relations. 

From the $2\times2$ covariance matrix $\Theta$,  we can write
\begin{align}
\epsilon_M & = M_{fit} \left( \frac{3 \, \epsilon_R}{R_{\Delta}} \right), \nonumber \\
\epsilon^2_R & = \theta_n^2 \; \Theta_{00} \; + \theta_a^2 \; \Theta_{11} \; + 2 \theta_n \theta_a \; \Theta_{10}  \nonumber \\
 \theta_n & = \frac{R_{\Delta} \ln(10)}{3-\epsilon} \nonumber \\
 \theta_a & = -\frac{\gamma -(1+0.5d) \tau}{3-\epsilon} R_{\Delta}^{4-\epsilon} \; \ln R_{\Delta} \;
   \ln\left(\frac{10^{\mathcal{A} -(1+0.5d) \mathcal{B}}}{R_0^{\gamma -(1+0.5d) \tau}}\right),
\end{align}
where $\theta_n$ and $\theta_a$ indicate the partial derivative of $R_{\Delta}$ with respect to the best-fit parameters.

We conclude this section by quoting some simple description of the radial dependence of the observed quantities $M_g$, $L$ and $T$.
By assuming that the distribution of the gas density is represented with a $\beta-$model,  $n_{\rm gas} \propto (1+x^2)^{-1.5 \beta}$, and the gas temperature profile with a functional form as in Vikhlinin et al. (2006; see also Baldi et al. 2012), and making the further assumption that $R_{500}$ is equal to 5 times the core radius $r_c = r/x$, we measure in the range $3 \leq x \leq 7$ the following radial behaviour
\begin{align}
M_g = M_{g, 0}\;  r^{2.73 -2.07 \beta} \nonumber \\
L = L_0 \; r^{1.17 -1.30 \beta} \nonumber \\
T = T_0 \; r^{-0.41 +0.13 \beta}.
\end{align}

\section{Catalogs of Mass estimates}

\begin{table*}
\caption{Best-fit results on the reconstructed masses of the objects in the samples from Mahdavi et al. (2013; M13), Maughan (2012; M12), Pratt et al. (2009; P09)
and Sun et al. (2009; S09) using the $M_g T$ gSR. The column ``CC'' indicates if the cluster hosts (1) or not (0) a cooling core (see Sect.~\ref{sect:samples} for details).
The redshifts quoted in the original work are used. A cosmology of $H_0=70$ km s$^{-1}$ Mpc$^{-1}$ and $\Omega_{\rm m} = 1 - \Omega_{\Lambda}=0.3$ is adopted.}
\begin{tabular}{cccccccccc} \hline
Cluster & Sample & $z$ & CC & $M_{\rm HSE}$ & $M_{fit, M13}$ & $M_{fit, M13-CC}$ & $M_{fit, M12}$ & $M_{fit, S09}$ \\ \hline
NGC1550 & S09 & 0.0124 & 1 & $0.33 \pm 0.04$ & $0.34 \pm 0.01$ & $0.38 \pm 0.04$ & $0.38 \pm 0.02$ & $0.37 \pm 0.02$ \\
MKW4 & M12 & 0.0199 & 1 & $0.79 \pm 0.10$ & $0.57 \pm 0.03$ & $0.66 \pm 0.08$ & $0.65 \pm 0.04$ & $0.61 \pm 0.02$ \\
MKW4 & S09 & 0.0200 & 1 & $0.51 \pm 0.07$ & $0.54 \pm 0.03$ & $0.63 \pm 0.08$ & $0.62 \pm 0.04$ & $0.57 \pm 0.02$ \\
3C442A & S09 & 0.0263 & 1 & $0.41 \pm 0.03$ & $0.39 \pm 0.03$ & $0.45 \pm 0.07$ & $0.45 \pm 0.04$ & $0.40 \pm 0.02$ \\
UGC5088 & S09 & 0.0274 & 1 & $0.15 \pm 0.03$ & $0.14 \pm 0.01$ & $0.18 \pm 0.04$ & $0.18 \pm 0.03$ & $0.14 \pm 0.02$ \\
NGC4104 & S09 & 0.0282 & 1 & $0.51 \pm 0.06$ & $0.45 \pm 0.03$ & $0.52 \pm 0.07$ & $0.51 \pm 0.04$ & $0.47 \pm 0.02$ \\
Abell1177 & S09 & 0.0316 & 1 & $0.55 \pm 0.08$ & $0.42 \pm 0.02$ & $0.49 \pm 0.07$ & $0.49 \pm 0.04$ & $0.45 \pm 0.02$ \\
NGC6269 & S09 & 0.0348 & 1 & $0.88 \pm 0.21$ & $0.70 \pm 0.03$ & $0.79 \pm 0.07$ & $0.78 \pm 0.03$ & $0.76 \pm 0.04$ \\
ESO306-017 & S09 & 0.0358 & 1 & $1.07 \pm 0.18$ & $1.02 \pm 0.05$ & $1.18 \pm 0.14$ & $1.17 \pm 0.07$ & $1.10 \pm 0.04$ \\
NGC5098 & S09 & 0.0368 & 1 & $0.21 \pm 0.04$ & $0.26 \pm 0.01$ & $0.30 \pm 0.04$ & $0.30 \pm 0.02$ & $0.28 \pm 0.01$ \\
MKW9 & M12 & 0.0382 & 1 & $0.88 \pm 0.20$ & $0.77 \pm 0.07$ & $0.93 \pm 0.18$ & $0.92 \pm 0.11$ & $0.79 \pm 0.07$ \\
Abell1983 & M12 & 0.0442 & 1 & $1.09 \pm 0.37$ & $0.78 \pm 0.05$ & $0.91 \pm 0.13$ & $0.90 \pm 0.07$ & $0.83 \pm 0.04$ \\
Abell160 & S09 & 0.0447 & 1 & $0.82 \pm 0.11$ & $0.69 \pm 0.02$ & $0.78 \pm 0.06$ & $0.77 \pm 0.03$ & $0.76 \pm 0.05$ \\
UGC842 & S09 & 0.0452 & 1 & $0.58 \pm 0.19$ & $0.47 \pm 0.03$ & $0.55 \pm 0.09$ & $0.54 \pm 0.05$ & $0.49 \pm 0.03$ \\
Abell2717 & M12 & 0.0498 & 1 & $1.10 \pm 0.12$ & $1.09 \pm 0.05$ & $1.25 \pm 0.14$ & $1.24 \pm 0.07$ & $1.17 \pm 0.05$ \\
Abell2717 & S09 & 0.0498 & 1 & $1.34 \pm 0.23$ & $1.12 \pm 0.05$ & $1.27 \pm 0.13$ & $1.26 \pm 0.06$ & $1.21 \pm 0.05$ \\
RXCJ1022+3830 & S09 & 0.0543 & 1 & $0.83 \pm 0.14$ & $0.75 \pm 0.04$ & $0.86 \pm 0.10$ & $0.85 \pm 0.05$ & $0.80 \pm 0.03$ \\
AS1101 & S09 & 0.0564 & 1 & $1.47 \pm 0.44$ & $1.44 \pm 0.03$ & $1.60 \pm 0.08$ & $1.58 \pm 0.05$ & $1.61 \pm 0.14$ \\
RXCJ2023.0-2056 & P09 & 0.0564 & 0 & $1.21 \pm 0.03$ & $1.15 \pm 0.05$ & $1.31 \pm 0.13$ & $1.30 \pm 0.06$ & $1.25 \pm 0.06$ \\
Abell133 & M12 & 0.0569 & 1 & $3.26 \pm 0.39$ & $2.58 \pm 0.08$ & $2.90 \pm 0.22$ & $2.88 \pm 0.10$ & $2.83 \pm 0.18$ \\
ESO351-021 & S09 & 0.0571 & 1 & $0.33 \pm 0.14$ & $0.32 \pm 0.02$ & $0.37 \pm 0.05$ & $0.37 \pm 0.03$ & $0.34 \pm 0.01$ \\
RXCJ2157.4-0747 & P09 & 0.0579 & 0 & $1.27 \pm 0.03$ & $1.16 \pm 0.03$ & $1.30 \pm 0.08$ & $1.28 \pm 0.04$ & $1.28 \pm 0.10$ \\
Abell3880 & S09 & 0.0581 & 1 & $1.55 \pm 0.44$ & $1.28 \pm 0.04$ & $1.44 \pm 0.10$ & $1.43 \pm 0.05$ & $1.42 \pm 0.10$ \\
Abell1991 & S09 & 0.0587 & 1 & $1.39 \pm 0.23$ & $1.34 \pm 0.05$ & $1.53 \pm 0.13$ & $1.51 \pm 0.06$ & $1.47 \pm 0.08$ \\
Abell1991 & M12 & 0.0592 & 1 & $1.27 \pm 0.17$ & $1.26 \pm 0.04$ & $1.42 \pm 0.12$ & $1.41 \pm 0.05$ & $1.38 \pm 0.08$ \\
RXCJ0345.7-4112 & P09 & 0.0603 & 1 & $0.98 \pm 0.02$ & $0.92 \pm 0.04$ & $1.05 \pm 0.11$ & $1.04 \pm 0.05$ & $0.99 \pm 0.04$ \\
RXCJ0225.1-2928 & P09 & 0.0604 & 0 & $1.01 \pm 0.04$ & $1.02 \pm 0.07$ & $1.21 \pm 0.19$ & $1.20 \pm 0.11$ & $1.07 \pm 0.06$ \\
Abell1795 & M12 & 0.0622 & 1 & $6.20 \pm 0.53$ & $5.20 \pm 0.10$ & $5.75 \pm 0.23$ & $5.69 \pm 0.17$ & $5.86 \pm 0.59$ \\
Abell1275 & S09 & 0.0637 & 1 & $0.72 \pm 0.24$ & $0.60 \pm 0.01$ & $0.67 \pm 0.04$ & $0.66 \pm 0.02$ & $0.67 \pm 0.05$ \\
Abell2092 & S09 & 0.0669 & 1 & $0.93 \pm 0.18$ & $0.69 \pm 0.02$ & $0.78 \pm 0.06$ & $0.77 \pm 0.03$ & $0.76 \pm 0.05$ \\
Abell2462 & S09 & 0.0733 & 1 & $0.91 \pm 0.13$ & $1.01 \pm 0.05$ & $1.16 \pm 0.12$ & $1.15 \pm 0.06$ & $1.09 \pm 0.04$ \\
Abell2029 & M12 & 0.0779 & 1 & $8.24 \pm 0.76$ & $8.31 \pm 0.17$ & $9.19 \pm 0.40$ & $9.11 \pm 0.27$ & $9.34 \pm 0.90$ \\
RXCJ1236.7-3354 & P09 & 0.0796 & 0 & $1.31 \pm 0.02$ & $1.24 \pm 0.05$ & $1.41 \pm 0.13$ & $1.40 \pm 0.06$ & $1.35 \pm 0.07$ \\
RXCJ2129.8-5048 & P09 & 0.0796 & 0 & $2.24 \pm 0.06$ & $2.19 \pm 0.07$ & $2.47 \pm 0.19$ & $2.44 \pm 0.09$ & $2.40 \pm 0.15$ \\
RXCJ1159+5531 & S09 & 0.0808 & 1 & $0.86 \pm 0.22$ & $0.67 \pm 0.04$ & $0.78 \pm 0.10$ & $0.77 \pm 0.05$ & $0.72 \pm 0.03$ \\
RXCJ0821.8+0112 & P09 & 0.0822 & 0 & $1.33 \pm 0.04$ & $1.22 \pm 0.04$ & $1.37 \pm 0.10$ & $1.36 \pm 0.04$ & $1.35 \pm 0.09$ \\
RXCJ1302.8-0230 & P09 & 0.0847 & 1 & $1.84 \pm 0.03$ & $1.70 \pm 0.04$ & $1.89 \pm 0.10$ & $1.88 \pm 0.05$ & $1.90 \pm 0.17$ \\
Abell1692 & S09 & 0.0848 & 1 & $1.01 \pm 0.25$ & $1.12 \pm 0.06$ & $1.30 \pm 0.17$ & $1.28 \pm 0.09$ & $1.19 \pm 0.05$ \\
Abell2597 & M12 & 0.0852 & 1 & $2.22 \pm 0.22$ & $2.16 \pm 0.06$ & $2.41 \pm 0.15$ & $2.39 \pm 0.07$ & $2.39 \pm 0.18$ \\
Abell478 & M12 & 0.0881 & 1 & $7.90 \pm 1.04$ & $7.61 \pm 0.15$ & $8.40 \pm 0.34$ & $8.32 \pm 0.25$ & $8.58 \pm 0.87$ \\
RXCJ0003.8+0203 & P09 & 0.0924 & 0 & $2.09 \pm 0.04$ & $2.07 \pm 0.08$ & $2.35 \pm 0.21$ & $2.33 \pm 0.10$ & $2.26 \pm 0.11$ \\
RXCJ2319.6-7313 & P09 & 0.0984 & 1 & $1.53 \pm 0.03$ & $1.37 \pm 0.03$ & $1.52 \pm 0.06$ & $1.50 \pm 0.04$ & $1.54 \pm 0.15$ \\
RXCJ0211.4-4017 & P09 & 0.1008 & 0 & $1.01 \pm 0.02$ & $0.91 \pm 0.03$ & $1.02 \pm 0.07$ & $1.01 \pm 0.03$ & $1.00 \pm 0.07$ \\
PKS0745-191 & M12 & 0.1028 & 1 & $7.27 \pm 0.75$ & $7.75 \pm 0.14$ & $8.50 \pm 0.26$ & $8.42 \pm 0.30$ & $8.82 \pm 1.02$ \\
RXCJ0049.4-2931 & P09 & 0.1084 & 0 & $1.66 \pm 0.05$ & $1.54 \pm 0.04$ & $1.72 \pm 0.10$ & $1.70 \pm 0.05$ & $1.71 \pm 0.14$ \\
RXCJ0006.0-3443 & P09 & 0.1147 & 0 & $3.78 \pm 0.12$ & $3.60 \pm 0.06$ & $3.95 \pm 0.13$ & $3.92 \pm 0.14$ & $4.09 \pm 0.47$ \\
RXCJ0616.8-4748 & P09 & 0.1164 & 0 & $2.64 \pm 0.05$ & $2.62 \pm 0.08$ & $2.95 \pm 0.23$ & $2.92 \pm 0.10$ & $2.88 \pm 0.18$ \\
RXCJ0145.0-5300 & P09 & 0.1168 & 0 & $4.11 \pm 0.08$ & $4.18 \pm 0.11$ & $4.67 \pm 0.29$ & $4.63 \pm 0.14$ & $4.63 \pm 0.35$ \\
RXCJ1516.3+0005 & P09 & 0.1181 & 0 & $3.09 \pm 0.04$ & $2.96 \pm 0.06$ & $3.28 \pm 0.14$ & $3.25 \pm 0.10$ & $3.33 \pm 0.32$ \\
RXCJ2149.1-3041 & P09 & 0.1184 & 1 & $2.22 \pm 0.05$ & $2.09 \pm 0.05$ & $2.33 \pm 0.12$ & $2.31 \pm 0.07$ & $2.34 \pm 0.20$ \\
RXCJ1141.4-1216 & P09 & 0.1195 & 1 & $2.21 \pm 0.02$ & $2.09 \pm 0.05$ & $2.33 \pm 0.12$ & $2.30 \pm 0.07$ & $2.33 \pm 0.20$ \\
RXCJ1516.5-0056 & P09 & 0.1198 & 0 & $2.54 \pm 0.05$ & $2.32 \pm 0.04$ & $2.54 \pm 0.07$ & $2.52 \pm 0.10$ & $2.65 \pm 0.32$ \\
Abell2550 & S09 & 0.1220 & 1 & $0.82 \pm 0.20$ & $0.80 \pm 0.03$ & $0.91 \pm 0.08$ & $0.90 \pm 0.04$ & $0.87 \pm 0.04$ \\
RXCJ1044.5-0704 & P09 & 0.1342 & 1 & $2.62 \pm 0.03$ & $2.41 \pm 0.04$ & $2.64 \pm 0.08$ & $2.61 \pm 0.10$ & $2.74 \pm 0.32$ \\
Abell1068 & M12 & 0.1375 & 1 & $3.87 \pm 0.28$ & $3.11 \pm 0.07$ & $3.46 \pm 0.19$ & $3.43 \pm 0.10$ & $3.47 \pm 0.30$ \\
RXCJ0605.8-3518 & P09 & 0.1392 & 1 & $3.73 \pm 0.06$ & $3.63 \pm 0.07$ & $4.01 \pm 0.17$ & $3.97 \pm 0.12$ & $4.08 \pm 0.40$ \\
RXCJ0020.7-2542 & P09 & 0.1410 & 0 & $3.73 \pm 0.06$ & $3.77 \pm 0.11$ & $4.23 \pm 0.28$ & $4.19 \pm 0.13$ & $4.17 \pm 0.30$ \\
RXCJ2218.6-3853 & P09 & 0.1411 & 0 & $4.71 \pm 0.11$ & $4.75 \pm 0.11$ & $5.28 \pm 0.27$ & $5.23 \pm 0.15$ & $5.32 \pm 0.47$ \\
Abell1413 & M12 & 0.1429 & 1 & $7.79 \pm 0.78$ & $6.66 \pm 0.13$ & $7.35 \pm 0.29$ & $7.28 \pm 0.23$ & $7.51 \pm 0.77$ \\
RXCJ2048.1-1750 & P09 & 0.1475 & 0 & $4.11 \pm 0.07$ & $3.83 \pm 0.08$ & $4.14 \pm 0.11$ & $4.10 \pm 0.20$ & $4.41 \pm 0.62$ \\
\hline \end{tabular}

\label{tab:mfit}
\end{table*}

\begin{table*}
\caption{Continue}
\begin{tabular}{cccccccccc} \hline
Cluster & Sample & $z$ & CC & $M_{\rm HSE}$ & $M_{fit, M13}$ & $M_{fit, M13-CC}$ & $M_{fit, M12}$ & $M_{fit, S09}$ \\ \hline
RXCJ0547.6-3152 & P09 & 0.1483 & 0 & $4.79 \pm 0.08$ & $4.73 \pm 0.09$ & $5.22 \pm 0.19$ & $5.17 \pm 0.16$ & $5.35 \pm 0.56$ \\
RXCJ2217.7-3543 & P09 & 0.1486 & 0 & $3.52 \pm 0.05$ & $3.35 \pm 0.06$ & $3.68 \pm 0.12$ & $3.65 \pm 0.13$ & $3.80 \pm 0.43$ \\
RXCJ2234.5-3744 & P09 & 0.1510 & 0 & $6.97 \pm 0.09$ & $6.90 \pm 0.13$ & $7.51 \pm 0.20$ & $7.44 \pm 0.33$ & $7.91 \pm 1.03$ \\
Abell2204 & M13 & 0.1520 & 1 & $8.70 \pm 0.60$ & $7.44 \pm 0.16$ & $8.04 \pm 0.23$ & $7.96 \pm 0.42$ & $8.60 \pm 1.25$ \\
Abell2204 & M12 & 0.1523 & 1 & $8.39 \pm 0.81$ & $7.84 \pm 0.14$ & $8.62 \pm 0.30$ & $8.53 \pm 0.29$ & $8.88 \pm 0.97$ \\
Abell2104 & M13 & 0.1530 & 0 & $5.80 \pm 0.80$ & $5.06 \pm 0.09$ & $5.56 \pm 0.19$ & $5.51 \pm 0.18$ & $5.73 \pm 0.63$ \\
RXCJ2014.8-2430 & P09 & 0.1538 & 1 & $5.10 \pm 0.06$ & $4.94 \pm 0.09$ & $5.39 \pm 0.15$ & $5.34 \pm 0.22$ & $5.65 \pm 0.71$ \\
Abell2259 & M13 & 0.1640 & 0 & $4.10 \pm 0.90$ & $4.04 \pm 0.09$ & $4.48 \pm 0.22$ & $4.44 \pm 0.13$ & $4.52 \pm 0.41$ \\
RXCJ0645.4-5413 & P09 & 0.1644 & 0 & $7.01 \pm 0.14$ & $6.93 \pm 0.13$ & $7.53 \pm 0.20$ & $7.46 \pm 0.34$ & $7.96 \pm 1.06$ \\
RXCJ0958.3-1103 & P09 & 0.1669 & 1 & $4.19 \pm 0.22$ & $4.32 \pm 0.13$ & $4.86 \pm 0.35$ & $4.82 \pm 0.16$ & $4.76 \pm 0.32$ \\
Abell1914 & M13 & 0.1710 & 0 & $9.20 \pm 0.90$ & $8.76 \pm 0.25$ & $9.83 \pm 0.67$ & $9.74 \pm 0.31$ & $9.68 \pm 0.68$ \\
Abell586 & M13 & 0.1710 & 0 & $3.90 \pm 0.60$ & $4.57 \pm 0.08$ & $4.99 \pm 0.14$ & $4.94 \pm 0.20$ & $5.22 \pm 0.64$ \\
MS0906.5+1110 & M13 & 0.1740 & 0 & $3.50 \pm 0.50$ & $5.24 \pm 0.13$ & $5.62 \pm 0.21$ & $5.57 \pm 0.35$ & $6.12 \pm 0.98$ \\
Abell2218 & M13 & 0.1760 & 0 & $4.30 \pm 0.60$ & $5.83 \pm 0.13$ & $6.47 \pm 0.32$ & $6.41 \pm 0.18$ & $6.52 \pm 0.59$ \\
Abell1689 & M13 & 0.1830 & 0 & $9.70 \pm 0.60$ & $9.50 \pm 0.17$ & $10.46 \pm 0.37$ & $10.36 \pm 0.34$ & $10.75 \pm 1.16$ \\
RXCJ1311.4-0120 & P09 & 0.1832 & 1 & $7.83 \pm 0.08$ & $8.11 \pm 0.15$ & $8.94 \pm 0.34$ & $8.86 \pm 0.28$ & $9.16 \pm 0.96$ \\
Abell383 & M13 & 0.1870 & 1 & $4.60 \pm 0.60$ & $2.77 \pm 0.05$ & $3.02 \pm 0.09$ & $2.99 \pm 0.12$ & $3.15 \pm 0.38$ \\
Abell383 & M12 & 0.1883 & 1 & $3.15 \pm 0.32$ & $3.29 \pm 0.07$ & $3.66 \pm 0.18$ & $3.62 \pm 0.10$ & $3.69 \pm 0.33$ \\
MS0440.5+0204 & M13 & 0.1900 & 1 & $2.80 \pm 0.50$ & $2.00 \pm 0.05$ & $2.23 \pm 0.12$ & $2.21 \pm 0.06$ & $2.23 \pm 0.19$ \\
Abell115S & M13 & 0.1970 & 0 & $4.20 \pm 0.30$ & $5.02 \pm 0.11$ & $5.42 \pm 0.17$ & $5.36 \pm 0.30$ & $5.83 \pm 0.88$ \\
Abell115N & M13 & 0.1970 & 1 & $4.10 \pm 0.20$ & $4.24 \pm 0.07$ & $4.64 \pm 0.13$ & $4.59 \pm 0.18$ & $4.84 \pm 0.59$ \\
Abell520 & M13 & 0.1990 & 0 & $7.30 \pm 0.30$ & $6.99 \pm 0.16$ & $7.78 \pm 0.41$ & $7.70 \pm 0.22$ & $7.80 \pm 0.67$ \\
Abell2163 & M13 & 0.2030 & 0 & $12.00 \pm 1.20$ & $14.36 \pm 0.34$ & $15.45 \pm 0.52$ & $15.29 \pm 0.90$ & $16.70 \pm 2.59$ \\
Abell963 & M13 & 0.2060 & 1 & $4.70 \pm 0.50$ & $4.83 \pm 0.12$ & $5.39 \pm 0.32$ & $5.34 \pm 0.15$ & $5.37 \pm 0.43$ \\
Abell209 & M13 & 0.2060 & 0 & $5.60 \pm 1.10$ & $6.85 \pm 0.13$ & $7.44 \pm 0.19$ & $7.37 \pm 0.34$ & $7.87 \pm 1.05$ \\
Abell222 & M13 & 0.2070 & 0 & $2.40 \pm 0.60$ & $3.46 \pm 0.10$ & $3.69 \pm 0.18$ & $3.65 \pm 0.26$ & $4.07 \pm 0.71$ \\
Abell223S & M13 & 0.2070 & 0 & $3.30 \pm 1.60$ & $4.76 \pm 0.08$ & $5.19 \pm 0.14$ & $5.14 \pm 0.21$ & $5.43 \pm 0.68$ \\
Abell1763 & M13 & 0.2230 & 0 & $3.90 \pm 0.70$ & $7.63 \pm 0.22$ & $8.13 \pm 0.39$ & $8.05 \pm 0.57$ & $8.98 \pm 1.56$ \\
Abell1942 & M13 & 0.2240 & 0 & $2.70 \pm 0.60$ & $3.19 \pm 0.06$ & $3.50 \pm 0.11$ & $3.47 \pm 0.12$ & $3.63 \pm 0.42$ \\
Abell2261 & M13 & 0.2240 & 1 & $6.60 \pm 1.00$ & $7.31 \pm 0.30$ & $7.68 \pm 0.57$ & $7.60 \pm 0.71$ & $8.76 \pm 1.80$ \\
Abell2219 & M13 & 0.2260 & 0 & $7.10 \pm 0.90$ & $10.26 \pm 0.23$ & $11.05 \pm 0.36$ & $10.94 \pm 0.63$ & $11.92 \pm 1.82$ \\
Abell2390 & M13 & 0.2280 & 1 & $11.00 \pm 0.90$ & $9.70 \pm 0.19$ & $10.51 \pm 0.28$ & $10.41 \pm 0.51$ & $11.19 \pm 1.56$ \\
Abell2111 & M13 & 0.2290 & 0 & $7.30 \pm 2.50$ & $5.20 \pm 0.09$ & $5.68 \pm 0.16$ & $5.62 \pm 0.22$ & $5.93 \pm 0.73$ \\
Abell2390 & M12 & 0.2302 & 1 & $11.05 \pm 1.11$ & $9.90 \pm 0.24$ & $10.63 \pm 0.38$ & $10.52 \pm 0.64$ & $11.53 \pm 1.82$ \\
Abell267 & M13 & 0.2310 & 0 & $5.70 \pm 0.60$ & $5.42 \pm 0.14$ & $6.06 \pm 0.37$ & $6.00 \pm 0.18$ & $6.01 \pm 0.46$ \\
MS1231.3+1542 & M13 & 0.2330 & 0 & $1.40 \pm 0.10$ & $2.05 \pm 0.16$ & $2.44 \pm 0.44$ & $2.43 \pm 0.26$ & $2.11 \pm 0.17$ \\
Abell1835 & M13 & 0.2530 & 1 & $9.90 \pm 0.70$ & $7.24 \pm 0.18$ & $7.77 \pm 0.30$ & $7.69 \pm 0.49$ & $8.46 \pm 1.37$ \\
Abell521 & M13 & 0.2530 & 0 & $5.00 \pm 1.30$ & $5.90 \pm 0.19$ & $6.28 \pm 0.33$ & $6.21 \pm 0.47$ & $6.97 \pm 1.25$ \\
Abell68 & M13 & 0.2550 & 0 & $5.10 \pm 1.00$ & $5.84 \pm 0.11$ & $6.44 \pm 0.24$ & $6.38 \pm 0.20$ & $6.60 \pm 0.69$ \\
MS1455.0+2232 & M13 & 0.2580 & 1 & $3.10 \pm 0.20$ & $3.49 \pm 0.08$ & $3.77 \pm 0.12$ & $3.73 \pm 0.21$ & $4.06 \pm 0.62$ \\
Abell1758W & M13 & 0.2790 & 0 & $11.50 \pm 1.60$ & $8.72 \pm 0.29$ & $9.85 \pm 0.79$ & $9.76 \pm 0.36$ & $9.55 \pm 0.56$ \\
Abell1758E & M13 & 0.2790 & 0 & $9.40 \pm 0.60$ & $9.53 \pm 0.19$ & $10.53 \pm 0.43$ & $10.43 \pm 0.31$ & $10.73 \pm 1.07$ \\
Abell697 & M13 & 0.2820 & 0 & $10.90 \pm 1.50$ & $10.90 \pm 0.19$ & $11.90 \pm 0.32$ & $11.78 \pm 0.48$ & $12.45 \pm 1.55$ \\
Abell959 & M13 & 0.2860 & 0 & $5.60 \pm 0.50$ & $5.50 \pm 0.10$ & $6.04 \pm 0.20$ & $5.98 \pm 0.21$ & $6.24 \pm 0.70$ \\
Abell611 & M13 & 0.2880 & 1 & $6.00 \pm 0.90$ & $5.56 \pm 0.14$ & $6.21 \pm 0.36$ & $6.15 \pm 0.18$ & $6.19 \pm 0.50$ \\
Abell2537 & M13 & 0.2950 & 0 & $5.90 \pm 0.90$ & $6.04 \pm 0.11$ & $6.60 \pm 0.18$ & $6.54 \pm 0.26$ & $6.90 \pm 0.85$ \\
MS1008.1-1224 & M13 & 0.3010 & 0 & $7.30 \pm 3.10$ & $4.45 \pm 0.09$ & $4.91 \pm 0.19$ & $4.87 \pm 0.15$ & $5.02 \pm 0.51$ \\
MS1358.1+6245 & M13 & 0.3280 & 1 & $7.60 \pm 0.90$ & $5.25 \pm 0.11$ & $5.81 \pm 0.25$ & $5.75 \pm 0.17$ & $5.90 \pm 0.57$ \\
MS1512.4+3647 & M13 & 0.3720 & 1 & $2.10 \pm 0.70$ & $2.01 \pm 0.05$ & $2.16 \pm 0.09$ & $2.13 \pm 0.14$ & $2.35 \pm 0.39$ \\
Abell370 & M13 & 0.3750 & 0 & $8.60 \pm 6.00$ & $6.75 \pm 0.13$ & $7.34 \pm 0.19$ & $7.26 \pm 0.33$ & $7.75 \pm 1.03$ \\
CL0024.0+1652 & M13 & 0.3900 & 1 & $3.10 \pm 4.70$ & $3.18 \pm 0.06$ & $3.47 \pm 0.10$ & $3.44 \pm 0.13$ & $3.62 \pm 0.44$ \\
Abell851 & M13 & 0.4070 & 0 & $7.40 \pm 2.30$ & $5.25 \pm 0.18$ & $5.56 \pm 0.32$ & $5.50 \pm 0.44$ & $6.22 \pm 1.16$ \\
MS1621.5+2640 & M13 & 0.4260 & 0 & $5.40 \pm 0.80$ & $5.48 \pm 0.11$ & $5.94 \pm 0.16$ & $5.88 \pm 0.28$ & $6.31 \pm 0.87$ \\
MACSJ0913.7+405 & M13 & 0.4420 & 1 & $4.80 \pm 0.70$ & $4.21 \pm 0.09$ & $4.67 \pm 0.21$ & $4.63 \pm 0.13$ & $4.73 \pm 0.44$ \\
RXJ1347.5-1145 & M13 & 0.4510 & 1 & $13.10 \pm 1.80$ & $12.39 \pm 0.23$ & $13.66 \pm 0.52$ & $13.53 \pm 0.43$ & $13.99 \pm 1.46$ \\
3C295 & M13 & 0.4640 & 1 & $3.90 \pm 1.00$ & $4.40 \pm 0.08$ & $4.81 \pm 0.14$ & $4.77 \pm 0.18$ & $5.02 \pm 0.61$ \\
RXJ1524.6+0957 & M13 & 0.5200 & 0 & $2.70 \pm 0.40$ & $3.20 \pm 0.06$ & $3.54 \pm 0.15$ & $3.50 \pm 0.10$ & $3.60 \pm 0.35$ \\
MS0015.9+1609 & M13 & 0.5410 & 0 & $13.40 \pm 1.90$ & $9.83 \pm 0.43$ & $10.29 \pm 0.82$ & $10.18 \pm 0.99$ & $11.81 \pm 2.50$ \\
MACSJ0717.5+374 & M13 & 0.5480 & 0 & $12.30 \pm 1.90$ & $13.15 \pm 0.41$ & $13.99 \pm 0.71$ & $13.85 \pm 1.03$ & $15.51 \pm 2.77$ \\
MS0451.6-0305 & M13 & 0.5500 & 0 & $7.80 \pm 1.00$ & $8.52 \pm 0.20$ & $9.48 \pm 0.51$ & $9.40 \pm 0.27$ & $9.50 \pm 0.81$ \\
\hline \end{tabular}

\end{table*}


\end{document}